\documentclass[conference]{templates/IEEEtran}

\usepackage[linesnumbered,ruled,algoruled,boxed,lined]{algorithm2e}
\usepackage{algorithmic}
\usepackage{tikz}
\usetikzlibrary{positioning,shapes}
\usepackage{amsmath}
\usepackage{listings}
\usepackage{tabularx}
\usepackage{longtable}
\usepackage{paralist}
\usepackage{balance}
\usepackage{xspace}
\usepackage{adjustbox}
\usepackage{multirow}
\usepackage{footnote}
\usepackage{xurl}
\usepackage{booktabs}
\usepackage{nicematrix}
\usepackage{wasysym}
\usepackage{float}
\usepackage{subcaption}
\usepackage{xcolor,colortbl}
\usepackage[group-separator={,},binary-units,per-mode=symbol,group-minimum-digits=4]{siunitx}

\usepackage{subcaption}
\definecolor{Gray}{gray}{0.85}
\newcolumntype{g}{>{\columncolor{Gray}}r}
\newcolumntype{h}{>{\columncolor{Gray}}c}

\newcommand{\cf}{cf.\@\xspace}
\newcommand{\ie}{i.\@\,e.\@\xspace}
\newcommand{\eg}{e.\@\,g.\@\xspace}

\newcommand{\etal}{et~al.\@\xspace}

\begin{document}
\title{Hope of Delivery: Extracting User Locations\\From Mobile Instant Messengers}

%\iffalse
\author{
\vspace{-1.2cm}
\emph{}\\
\emph{This is an extended version of the original paper published at}\\
\emph{The Network and Distributed System Security (NDSS) Symposium 2023}\\
\vspace{0.5cm}
https://dx.doi.org/10.14722/ndss.2023.23188\\

\IEEEauthorblockN{
	Theodor Schnitzler\IEEEauthorrefmark{1}\IEEEauthorrefmark{2},
	Katharina Kohls\IEEEauthorrefmark{3}, 
	Evangelos Bitsikas\IEEEauthorrefmark{4}\IEEEauthorrefmark{5}, and
	Christina Pöpper\IEEEauthorrefmark{5}}
\IEEEauthorblockA{
	\IEEEauthorrefmark{1}Research Center Trustworthy Data Science and Security, TU Dortmund, Germany \hskip 0.15cm \IEEEauthorrefmark{2}Ruhr-Universität Bochum, Germany \\ \IEEEauthorrefmark{3}Radboud University, Netherlands \hskip 0.1cm \IEEEauthorrefmark{4}Northeastern University, USA \hskip 0.1cm \IEEEauthorrefmark{5}New York University Abu Dhabi, UAE}
	
	theodor.schnitzler@tu-dortmund.de \hskip 0.2cm
	kkohls@cs.ru.nl \hskip 0.2cm
	bitsikas.e@northeastern.edu \hskip 0.2cm
	christina.poepper@nyu.edu
	
	\\
}
%\fi

\iffalse
\IEEEoverridecommandlockouts
\makeatletter\def\@IEEEpubidpullup{6.5\baselineskip}\makeatother
\IEEEpubid{\parbox{\columnwidth}{
    Network and Distributed System Security (NDSS) Symposium 2023\\
    27 February - 3 March 2023, San Diego, CA, USA \\
    ISBN 1-891562-83-5\\
    https://dx.doi.org/10.14722/ndss.2023.23188\\
    www.ndss-symposium.org
}
\hspace{\columnsep}\makebox[\columnwidth]{}}
\fi

\maketitle

\begin{abstract}
Mobile instant messengers such as WhatsApp use delivery status notifications in order to inform users if a sent message has successfully reached its destination.  This is useful and important information for the sender due to the often asynchronous use of the messenger service. However, as we demonstrate in this paper, this standard feature opens up a timing side channel with unexpected consequences for user location privacy. We investigate this threat conceptually and experimentally for three widely spread instant messengers. We validate that this information leak even exists in privacy-friendly messengers such as Signal and Threema.

Our results show that, after a training phase, a messenger user can distinguish different locations of the message receiver.
Our analyses involving multiple rounds of measurements and evaluations show that the timing side channel persists independent of distances between receiver locations~--~the attack works both for receivers in different countries as well as at small scale in one city.
For instance, out of three locations within the same city, the sender can determine the correct one with more than \SI{80}{\percent} accuracy.
Thus, messenger users can secretly spy on each others' whereabouts when sending instant messages.
As our countermeasure evaluation shows, messenger providers could effectively disable the timing side channel by randomly delaying delivery confirmations within the range of a few seconds.
For users themselves, the threat is harder to prevent since there is no option to turn off delivery confirmations.
\end{abstract}

\section{Introduction}
\label{sec:intro}

% OVERVIEW
In recent years, messaging applications (or messengers) have become the de-facto standard for mobile communication. They have transitioned into integral parts of daily lives, with the most prominent messenger, WhatsApp, connecting more than two billion monthly active users world-wide~\cite{statista21:most-popular-global}.
% USE CASES / EXAMPLES
Messengers are used in a wide range of scenarios, from informal communication among working colleagues~\cite{loch19:whatsapp-workplace} and social engagement among elderly people~\cite{miller21:global-smartphone-beyond}, to parents coordinating school matters~\cite{sebugwaawo18:whatsapp-most-preferred} and citizens organizing neighborhood watches~\cite{dalmeijer19:whatsapp-neighbourhood-watch}.
In some cases, messengers are also used for official communication with government authorities~\cite{kohler20:department-health-abudhabi,purz20:governments-worldwide-messaging}, thus composing large and heterogeneous sets of contacts in one application per user.

Whenever a user sends a message in a messenger, the client application displays the current status of the message~--~from being in transit, processed and forwarded by the messenger server, to delivered to the recipient, and (if enabled) read by the recipient~\cite{ariano20:what-check-marks}, often indicated by small symbols such as checkmarks.
This is helpful information for users to track if a message has successfully reached its destination.

% PROBLEM STATEMENT
However, as we will demonstrate in our paper, this feature can also serve as a side channel that allows to learn sensitive information about message recipients, such as revealing information about their current whereabouts, with undesired potential harm to location privacy. 

% WHAT WE DO
In more details, we conduct a series of experiments in Signal~\cite{whisper10:signal}, Threema~\cite{threema12:threema}, and WhatsApp~\cite{whatsapp09:whatsapp} to evaluate and demonstrate to what extent we can classify different message receivers and their respective locations based on delivery notification timings of a set of subsequently sent messages. 
Deriving sensitive information about someone by sending them a few messages is problematic because it is simple, rather unsuspicious, and hard to mitigate.
Users cannot effectively prevent receiving messages from people in their contact list, except for permanently blocking them and, therefore, stopping having mobile conversations with them at all.  

Based on characteristics such as the location of a receiver, delivering a message and returning the respective confirmation takes a specific amount of time. 
Physical transmissions on the Internet are influenced by the travelled distance, they depend on the network topology, \ie, routing and the hops in-between, and processing by the messaging service. 
We show that \emph{sending messages} using each of these three messengers \emph{to receivers at different locations results in different and distinguishable delivery notification timing patterns}.

This issue is critical for multiple reasons:  
First, all three messengers we examine are generally considered secure as they use end-to-end encryption between clients. It is not intuitive for users that the mere usage of the messenger service may leak information about their whereabouts. Second, Signal and Threema are best known for their focus on privacy~--~ Signal's protocol serves as the blueprint for provably secure key establishment between clients~\cite{cohn17:formal-analysis-signal} and has been adapted by other applications such as WhatsApp. Leaking information of the user's location contradicts this notion of privacy. 
Third, a user cannot do much about someone in their contact list sending them instant messages.
Other than read receipts that can be turned off by the receiver for privacy reasons, there is no such option for delivery notifications~\cite{whatsapp21:check-read-receipts}.

% SECURITY-FOCUS
In order to experimentally validate this concept we need to take into account the server infrastructures of messengers. This information is not publicly shared and it is a challenge in itself to reliably extract the relevant information such as the number and locations of messenger servers.  
To this end, we conduct experiments to collect and aggregate information about the geographical distribution of servers of popular instant messaging services and analyze if and how knowledge about the messaging server in use affects the outcome of the delivery timing evaluation. We note that the server infrastructure setup does not change frequently, so this step would not have to be redone for each user localization attempt. Beyond the proof-of-concept attack done in this work, knowledge about the messenger infrastructure may  %turn out to 
be useful for other purposes. 

In summary, our paper makes the following contributions: 
\begin{compactenum}
\item \textbf{Messenger Infrastructure Analysis.} 
We aggregate and provide an overview of the geographical distribution of servers of mobile messaging services from a series of experiments to discover and analyze their infrastructures.
\item \textbf{Empirical Messaging Experiments.}
We conduct large-scale measurements collecting the transmission timings of message delivery notifications between devices in multiple locations in Europe and the Middle East.
\item \textbf{Attack and Countermeasure Evaluation.}
We demonstrate to what extent we can distinguish different receivers and their respective locations from each other based on the measured delivery notification timings.
We also show that this threat can be mitigated by randomly delaying delivery notifications in the range of a few seconds.
\end{compactenum}

\subsubsection*{Experimental Overview}
Figure~\ref{fig:overview} provides an overview of our experiments for each of the three parts, their results and connections with each other.
We start with infrastructure discovery experiments that result in sets of server locations used to determine the infrastructure overhead in the messaging experiments.
At the core of our study, we use sequences of message delivery notification timings to classify receiver locations at different granularity levels and measure the accuracy.

\subsubsection*{Disclosure Process}
The timing side channel exploited in this paper may potentially affect the location privacy of millions of messenger users. 
Following the guidelines of responsible disclosure, we got in contact with the providers of the messenger apps (Signal, Threema, WhatsApp) and reported the vulnerability to them prior to the submission of this paper in May~2022.
Whereas Signal and WhatsApp have not acknowledged the issue to date (October~2022), we have exchanged ideas for mitigating the problem with Threema and they are currently evaluating how specific countermeasures (\cf~Section~\ref{sec:countermeasures}) would affect user experience.

\definecolor{redline}{RGB}{184,84,80}
\definecolor{redfill}{RGB}{248,206,204}

\pgfdeclarelayer{background}
\pgfsetlayers{background,main}
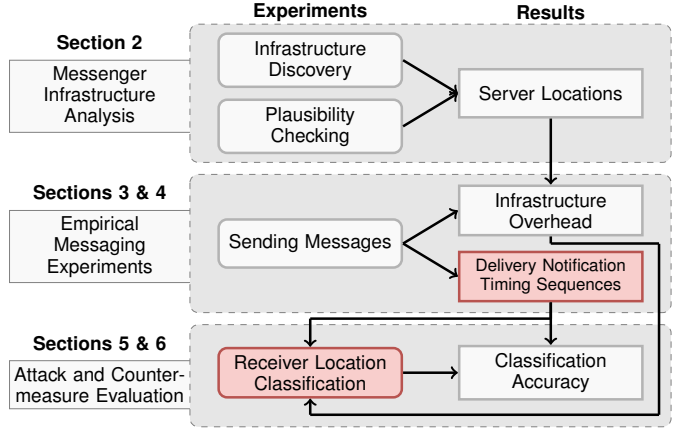
\begin{figure}[tb]
\centering
\begin{adjustbox}{width=\columnwidth}
\begin{tikzpicture}[
    sectionnode/.style={text width=3cm, text centered,font=\sffamily\small},
	squarednode/.style={rectangle, draw=gray!60, fill=gray!5, very thick, minimum size=8mm, text width=8em, text centered, font=\sffamily\small},
	roundednode/.style={rectangle, draw=gray!60, fill=gray!5, very thick, rounded corners, minimum size=8mm, font=\sffamily\small, text width=8em, text centered},
	roundednodered/.style={rectangle, draw=redline, fill=redfill, very thick, rounded corners, minimum size=8mm, font=\sffamily\small, text width=8em, text centered},
	attacknodered/.style={rectangle, draw=redline, fill=redfill, very thick,  minimum size=8mm, font=\sffamily\footnotesize, text width=8em, text centered},
	textnode/.style={minimum size=8mm, text width=6em, text centered},
	font=\sffamily
    ]
	%LEFT BLOCK
    \node[roundednode] (infradisc)  at (5,-0.6) {Infrastructure Discovery};
    \node[roundednode] (infracheck)   at (5,-1.7) {Plausibility\\Checking};
    \node[squarednode] (infraloc)   at (9,-1.15) {Server Locations};
	
	\node[roundednode] (msgdata)  at (5,-3.65) {Sending Messages};
    \node[attacknodered] (msgtimings)   at (9,-4.2) {Delivery Notification \\Timing Sequences};
    \node[squarednode] (msgoverhead)   at (9,-3.1) {Infrastructure Overhead};
	
    \node[roundednodered] (mlclassify) at (5,-5.8) {Receiver Location\\ Classification};
	\node[squarednode] (mlaccuracy) at (9,-5.8) {Classification\\ Accuracy};

    %Lines
	\draw[->, very thick] (infradisc.east) -- (infraloc.west);
	\draw[->, very thick] (infracheck.east) -- (infraloc.west);

	\draw[->, very thick] (infraloc.south) -- (msgoverhead.north);
	\draw[->, very thick] (msgdata.east) -- (msgtimings.west);
	\draw[->, very thick] (msgdata.east) -- (msgoverhead.west);

	\draw[-, very thick] (msgtimings.south) -- (9,-4.9);
	\draw[-, very thick] (9,-4.9) -- (5,-4.9);
	\draw[->, very thick] (5,-4.9) -- (mlclassify.north);
	\draw[->, very thick] (msgtimings.south) -- (mlaccuracy.north);
	
	\draw[->, very thick] (mlclassify.east) -- (mlaccuracy.west);
	
	% complex arrow from infra-overhead to classification
	\draw[-, very thick] (msgoverhead.south) -- (9,-3.65);
	\draw[-, very thick] (9,-3.65) -- (10.8,-3.65);
	\draw[-, very thick] (10.8,-3.655) -- (10.8,-6.5);
	\draw[-, very thick] (10.8,-6.5) -- (5,-6.5);
	\draw[->, very thick] (5,-6.5) -- (mlclassify.south);
	
    \node[sectionnode] at (5,0.2) {\textbf{Experiments}};
    \node[sectionnode] at (9,0.2) {\textbf{Results}};
    %\draw[-, dashed] (7,0.3) -- (7,-6.7);

    \begin{pgfonlayer}{background}

    % Section Mapping      

    %\path[fill=gray!5, draw=black!50]
	%	  (-3.4,-4.2) rectangle (-0.8,-4.8) ;
    \path[fill=gray!5, draw=black!50]
    (0,-0.6) rectangle (3.5,-1.8) ;	
    \node[sectionnode] at (1.5,-0.3) {\textbf{Section 2}};
    \node[sectionnode] at (1.5,-1.2) {Messenger\\ Infrastructure\\ Analysis};
    
    \path[fill=gray!5, draw=black!50]
    (0,-3.1) rectangle (3.5,-4.3) ;
    \node[sectionnode] at (1.5,-2.8) {\textbf{Sections 3 \& 4}};
	\node[sectionnode] at (1.5,-3.7) {Empirical\\ Messaging\\ Experiments};

    \path[fill=gray!5, draw=black!50]
	(0,-5.6) rectangle (3.5,-6.4) ;
	\node[sectionnode] at (1.5,-5.3) {\textbf{Sections 5 \& 6}};
	\node[sectionnode] at (1.5,-6) {Attack and Counter-\\ measure Evaluation};

	% BG Boxes left
    \path[fill=gray!20,rounded corners, draw=black!50, dashed]
		  (3,0) rectangle (11,-2.3) ;		  
	\path[fill=gray!20,rounded corners, draw=black!50, dashed]
		  (3,-2.5) rectangle (11,-4.8) ;
	\path[fill=gray!20,rounded corners, draw=black!50, dashed]
		  (3,-5) rectangle (11,-6.7) ;  
%	\path[fill=gray!20,rounded corners, draw=black!50, dashed]
%          (-10,-3.8) rectangle (-3.3,-5.2) ;

    \end{pgfonlayer}
\end{tikzpicture}
\end{adjustbox}
\caption{Structural overview of the sequence of experiments (rounded nodes) and their outcomes (square nodes) in our paper and how the three main parts build upon each other.}
\label{fig:overview}
\end{figure}

\section{Messenger Infrastructure Analysis}
\label{sec:messenger-infrastructures}

Our first goal is to obtain a comprehensive overview of the infrastructures of the messengers we use in our experiments, \ie, for Signal, Threema, and WhatsApp.
For the delivery notification timing analysis, knowledge about the infrastructure is crucial to assess the different parts of the connection between sender and receiver, their distances, and timings.

\subsection{Discovery and Aggregation}
\label{sec:infrastructure-discovery}
In order to gain first insights into the messenger infrastructures, we conduct a set of experiments to identify servers used by messaging services.
In the first step, we set up two smartphones running client applications for all messengers under consideration and capture their network traffic when the applications are running.
From the collected captures, we extract the IP addresses of the servers that the application on the smartphone connects to.
Since we assume that messenger servers are geographically distributed, the resulting sets of IP addresses may only represent specific fractions of the messenger infrastructures, \ie, they comprise servers near to our own location.

To broaden the perspective derived from our local observations, we perform a two-step DNS analysis, as follows:
\begin{compactenum}[(1)]
    \item For all IP addresses that appear in the communication using one of the messaging applications, we perform reverse DNS lookups to learn what (sub)domain names are used by the messenger operations.
    \item For each domain name in the set derived from reverse look-ups, we perform federated DNS resolving from multiple locations across all continents.
\end{compactenum}

\noindent
We continue to describe the exact procedures for each messenger individually.
\subsubsection{Signal}
For Signal, two specific IPv4 addresses are in use.
Reverse DNS lookups point to the same domain name operated by Amazon Web Services (AWS), also when we perform these lookups from different geographical locations.
When we resolve the resulting domain name, the same two IP addresses are returned, irrespective of the location. 
Even though the order of the two addresses varies, there is no indication that one address is preferred over the other at specific locations.

\subsubsection{Threema}
For Threema, we identify two similar IP addresses from the same IPv4/24 address range, for one of which the reverse DNS lookup points to a \texttt{threema.ch} domain name.
Reverse lookup fails for the other address.
We manually identify several more IP addresses whose domain names are resolved to \texttt{threema.ch}, resulting in an extended set of \num{12} IP addresses.
However, it is unclear if all these IP addresses are actually used for the messaging application of if they serve other purposes related to the same domain.

\subsubsection{WhatsApp}
Our reverse domain name resolving of server IP addresses reveals that WhatsApp establishes connections to servers in five different domain name ranges.
Additionally, different servers within the same domain name range have been used.
Irrespective of the location at which we perform the reverse DNS lookup for a particular IP address, it is resolved to the exact same domain name. Across the three messengers, we discover the largest number of different IP addresses when we explore the network traffic of WhatsApp.

The WhatsApp domain names within the same namespace only differ in 3-letter strings which appear to be IATA airport codes\footnote{\scriptsize\url{https://www.iata.org/en/publications/directories/code-search/?airport.search=}} near our experimental locations. 
Random checks of additional domain names with the identifier replaced with different ones (in other regions all over the world) reveal further IP addresses, strengthening our assumption.

Since all tested domain names resolve to similar IP addresses in five different IPv4/16 subnets, we conduct a full search of the respective address ranges.
We record all domain names and their corresponding IPv4 addresses that contain a reference to WhatsApp (\cf~Table~\ref{tab:wa-namespaces}).
We further extend the resulting set by manually spot-checking even more identifiers, which leads to a small number of additional servers.
In total, our set of discovered WhatsApp servers comprises \num{410} server instances using \num{143} different location identifiers.

\begin{table}[h]
\caption{Namespace prefixes used by WhatsApp servers.~\label{tab:wa-namespaces}}
\centering
\begin{tabular}{@{}lcc@{}}
\toprule
Namespace (Prefix) &  No.~of IPs & No.~of Locations\\
\midrule
fna-whatsapp & 126 & 75\\
whatsapp-chatd-edge & 94 & 73\\
whatsapp-chatd-msgr-edge & 92 & 72\\
whatsapp-cdn & 92 & 72\\
whatsapp-pp & 6 & 4\\
\midrule
Total Unique IPs/Locations & 410 & 143\\
\bottomrule
\end{tabular}
\end{table}

\subsection{Location Analysis}

In the next step, we map messenger servers to their individual geographical location and validate the mapping with the help of simple plausibility checks.
We initially map each messenger server identified in Section~\ref{sec:infrastructure-discovery}, \ie, their IP addresses, to a specific geographical location.
We use different strategies depending on the information that we can obtain per messenger.

Little official information about messenger infrastructures is made public by their providers.
In the set of messengers we explored, only Threema mentions that their servers are located in the Zurich area, Switzerland~\cite{threema21:security-privacy-faq}.
For Signal, no official information is available but several sources indicate that servers are hosted by AWS at the US east coast~\cite{bahramali20:practical-traffic-analysis,duckett21:aws-north-virginia,sharwood21:aws-us-east,williams21:secure-messaging-apps} which is presumably located near Ashburn, VA.

The only information we find with relation to WhatsApp is a list of the locations of Facebook data centers on their website~\cite{facebook21:data-centers}.
It is, however, unclear if these locations are also related to WhatsApp.
We additionally take into account the presumable IATA location identifiers within the domain names associated with IP addresses used by WhatsApp.
We perform look-ups for all \num{143} codes that appear in our data set and use the resulting city as baseline location for the server.
In a few cases, identifiers could not be resolved~--~and we manually annotate them. 
For example, the codes \emph{frx} and \emph{frt} most likely belong to the area of Frankfurt, Germany (whose original IATA identifier is \emph{fra}).

\begin{figure}[tb]
\centering
\includegraphics[width=\columnwidth]{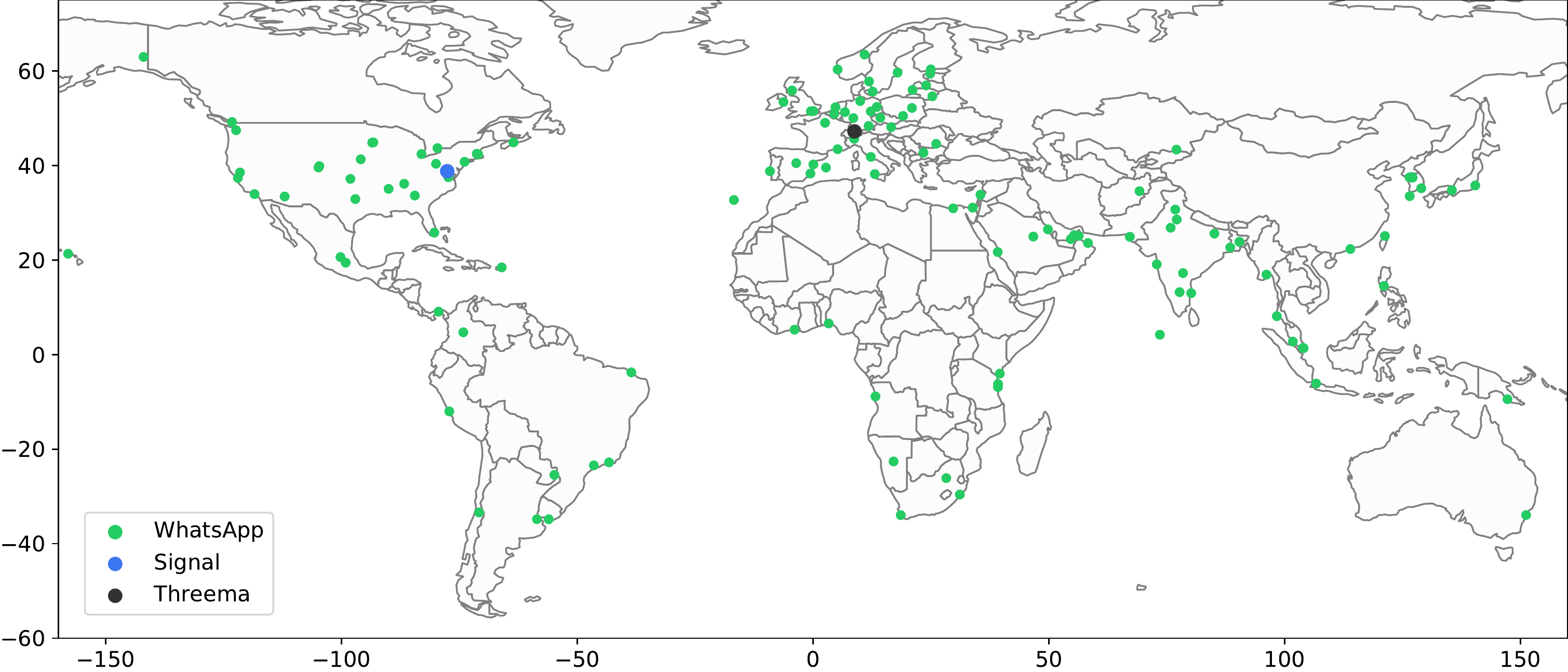}
\caption{Locations of Signal, Threema, and WhatsApp servers around the world (larger version in the Appendix).~\label{fig:worldmap_small}} 
\end{figure}

We continue with a series of systematic Ping and Traceroute experiments from different geographical locations using a public API provided by CheckHost~\cite{checkhost21:api-overview}.
Over a period of four weeks we collect ping and routing information to all messenger servers.
To confirm a location candidate as correct, we require that the shortest Ping time is received by the probe host that is closest to the location candidate and only accept minor deviations.

Whereas for WhatsApp and Threema the results are consistent and confirm our initial assumptions about the baseline, the case is more difficult for Signal. 
Ping information is heavily inconsistent with results being within less than \SI{10}{ms} from all different continents, which suggests that they are returned from different physical locations close to each of the probing hosts.
While Traceroute information can only be partially retrieved for Signal, they include traces with hosts that are likely located in the US, which again strengthens the initial assumption of Signal servers to be US-based.

Figure~\ref{fig:worldmap_small} shows our extracted geographical overview of the server locations for the three messengers (a larger/more readable version can be found in  Figure~\ref{fig:worldmap} in Appendix~\ref{sec:appendix-worldmap}).

\section{Message Status Timing Side Channel}
\label{sec:experiments}

The main idea of the attack we present is the use of a timing side channel provided by message status information to derive characteristics of a target user's Internet connection.
Whenever two users are in each other's contact list of a mobile messaging application, \ie, they have accepted to be in a conversation on that messenger, the application shows status information for exchanged messages.

Small icons (\eg, check marks) along with each message indicate whether a message has been sent to the messenger server, delivered to the receiver, or read by the receiver.
The messages between users as well as the information about the message status are exchanged through TCP messages between the client application and the messenger server.
We measure the time between sending a message (\ie, the TCP packets containing the message leaving the sender's device) and the server and delivery confirmations (\ie, the TCP packets containing these confirmations) arriving at the sender's device.
Observing the resulting timing difference allows us to reason about characteristics of the receiver, such as their location, or their network connection.
A schematic overview of the information flow is depicted in Figure~\ref{fig:info_flow}.

\definecolor{redline}{RGB}{184,84,80}
\definecolor{redfill}{RGB}{248,206,204}

\pgfdeclarelayer{background}
\pgfsetlayers{background,main}
\begin{figure}[tb]
\centering
\begin{adjustbox}{width=\columnwidth}
\begin{tikzpicture}[
    sectionnode/.style={text width=3cm, text centered,font=\sffamily\small},
	squarednode/.style={rectangle, draw=gray!60, fill=gray!5, very thick, minimum size=8mm, text width=8em, text centered, font=\sffamily\small},
	roundednode/.style={rectangle, draw=gray!60, fill=gray!5, very thick, rounded corners, minimum size=8mm, font=\sffamily, text width=10em, text centered},
	attacknodered/.style={rectangle, draw=redline, fill=redfill, very thick,  minimum size=8mm, font=\sffamily\footnotesize, text width=8em, text centered},
	textnode/.style={minimum size=8mm, text width=6em, text centered},
	font=\sffamily
    ]
	%LEFT BLOCK
    \node[roundednode] (sender)  at (0,0) {(S)ender};
    \node[roundednode] (server)   at (4,0) {(M)essenger Server};
    \node[roundednode] (receiver)   at (8,0) {(R)eceiver};
    
    \node[draw=none,fill=none] at (1,-1.1){\includegraphics{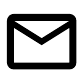}};
    \node[draw=none,fill=none] at (5,-1.2){\includegraphics{tikz/msg.pdf}};
    \node[draw=none,fill=none] at (3,-2.1){\includegraphics{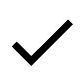}};
    \node[draw=none,fill=none] at (7,-3.4){\includegraphics{tikz/check1.pdf}};
    \node[draw=none,fill=none] at (3,-3.5){\includegraphics{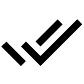}};
	\node[draw=none,fill=none] at (-0.6,-2.4){\includegraphics{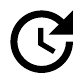}};
    %Lines
	\draw[-, very thick] (sender.south) -- (0,-4);
	\draw[-, very thick] (server.south) -- (4,-4);
	\draw[-, very thick] (receiver.south) -- (8,-4);
	
	\draw[->, very thick, dashed] (0,-1.5) -- (4,-1.5);
	\draw[->, very thick, dashed] (4,-1.6) -- (8,-1.6);
	\draw[->, very thick, dashed] (4,-1.7) -- (0,-1.7);
	
	\draw[->, very thick, dashed] (8,-3) -- (4,-3);
	\draw[->, very thick, dashed] (4,-3.1) -- (0,-3.1);

\end{tikzpicture}
\end{adjustbox}
\caption{Schematic overview of the message flow from the senders' perspective. The illustration is simplified since sender and receiver can be connected to different messenger servers.}
\label{fig:info_flow}
\end{figure}

Factors such as the travelled distance between sender, server, and receiver, routing through the Internet between these entities, as well as processing at the messenger server and at hops in-between can affect the observed timings.
Repeatedly sending messages to receivers under different conditions (\eg, location, network connection) and observing the timings between messages allows us to learn characteristics of the timings under these conditions in a controlled setup.
For different receiver locations, the duration or the distribution of RTTs may be different, \eg, longer times likely represent longer distances between the communication partners.

Within our experiments, we demonstrate to what extent it is feasible to determine certain receiver characteristics upon observing delivery notification timings.

\subsection{Threat Model}
From a \emph{technical} perspective, the adversary is required to operate a regular smartphone that is capable of running a messenger application. 
The adversary additionally needs to be able to access and analyze their own TCP traffic to extract timing information. 
This traffic can be captured either on a node in their local network, or directly running on the smartphone when running a packet capture app.

As an \emph{operational} requirement for the attack, adversary and victim must be in each other's contact lists in the messenger. 
Thus, the threat is limited to parties who likely know each other, as the attack can only be conducted against users who have added the adversary to their contacts.
However, the various contexts in which people have messenger conversations, be it in personal (extended family, acquaintances),  professional (\eg, work collaborators) or other contexts (\eg, interaction with public institutions, clubs, authorities, within neighborhoods) in combination with low technical requirements still yield a considerable threat scope within social circles, \eg, for stalking.

In an initial training phase, the adversary sends messages to the victim and learns timing characteristics while knowing their whereabouts. 
Subsequently, the adversary can send new messages to the victim, and determine their location or network connection out of the set of previously seen plausible ones.
Since the attack entails sending messages, the adversary's behavior might be observed by the victim and appear suspicious.
Therefore, the attacker might leverage timings of messages they send anyway which would, however, narrow down the practical threat scope to people who regularly exchange larger numbers of messages.

\subsection{Setup}
\label{sec:measurement-setup}

We conduct measurements while sending messages between multiple smartphones in different geographical locations. Our setup comprises two types of devices:

\begin{compactenum}[(i)]
\item \emph{Active} devices are used to  send messages to other devices. Each active device is connected via USB to a computer scheduling the experiment and controlling the smartphone via Android Debug Bridge (ADB).
\item \emph{Passive} devices are used to receive messages from active devices. The only requirement for a passive device is having an active Internet connection. 
\end{compactenum}

We conduct two rounds of measurements serving different purposes:
\paragraph*{1)} In the first round, we conduct long-distance measurements with devices distributed across different countries.
During this round of measurements, each device is assigned a specific, permanent location.
Out of three devices for active measurements, two are located in Germany (\emph{DE-11} and \emph{DE-12}) and one in Greece (\emph{GR-11}).
Our setup comprises three more passive devices, located in Germany (\emph{DE-13}), the Netherlands (\emph{NL-11}) and the Middle East (\emph{AE-11}).
This experiment is meant to demonstrate a proof of concept that the message-status timing side channel actually exists. For the sake of simplicity, all devices operated on a WiFi Internet connection for these measurements.
\paragraph*{2)} In a second round of measurements, we send messages from a single active device to passive devices at locations closer to each other, \ie, within the same city, and also rotate passive devices through these locations.
Furthermore, passive devices switch between WiFi and cellular Internet connections.
We replicate this type of setup in Germany~(\emph{DE-2X}) and the Middle East~(\emph{AE-2X}).
This round of measurements is meant to demonstrate a more practical and realistic attack scenario, imitating a natural everyday behavior of a target messenger client, \eg, being at their home and work location (WiFi) and moving in between and around (cellular).
Furthermore, this second round also shows to what extent the attack works at a smaller scale, which is less obvious than comparing timings at country level.
%\end{compactenum}

\begin{table}[tbp]
\centering
\caption{Devices and locations in our measurements.~\label{tab:devices-overview}}
\resizebox{\columnwidth}{!}{
\begin{tabular}{@{}l@{\hspace{0.15cm}}l@{\hspace{0.05cm}}c@{\hspace{0.15cm}}l@{}}
\toprule
ID & Model (Year) & Type & Locations\\
\midrule
\multicolumn{4}{l}{\emph{Round 1}}\\
AE-11 & Huawei P40 (2020) & P & AE-A (W)\\
DE-11 & Xiaomi Mi A3 (2019) & A,P & DE-A (W)\\
DE-12 & Huawei P8 Lite (2017) & A,P & DE-B (W)\\
DE-13 & OnePlus 7 Pro (2019) & P & DE-B (W)\\
GR-11 & Samsung Note 10+ (2019) & A,P & GR-A (W)\\
NL-11 & Samsung S6 (2015) & P & NL-A (W)\\
\midrule
\multicolumn{4}{l}{\emph{Round 2 (United Arab Emirates)}}\\
AE-21 & Huawei P40 (2020) & A & AE-B (W)\\
AE-22 & Samsung Note 10 (2019) & P & AE-A, AE-D (W, 4G+)\\
AE-23 & Samsung S22 (2022) & P & AE-B (W, 5G) \\
AE-24 & Nokia X10 (2021) & P & AE-C (W, 4G+)\\
\midrule
\multicolumn{4}{l}{\emph{Round 2 (Germany)}}\\
DE-21 & Huawei P8 Lite (2017) & A & DE-A (W)\\
DE-22 & Huawei P8 Lite (2017) & P & DE-A (W), DE-B (W, 4G), DE-C (W)\\
DE-23 & Google Pixel 3a (2019) & P & DE-A (W, 4G), DE-B (W), DE-D (W)\\
DE-24 & Samsung S6 (2015) & P & DE-A (W), DE-B (W, 4G),  DE-E (W)\\
\midrule
\multicolumn{4}{@{}l@{}}{Locations: \emph{AE-A,B,C,D:} Abu Dhabi, UAE; \emph{DE-A,B,D,E:} Bochum, Germany;}\\
\multicolumn{4}{@{}l@{}}{\emph{DE-C:} Essen, Germany; \emph{NL-A:} Nijmegen, Netherlands; \emph{GR-A:} Athens, Greece}\\
\bottomrule
\end{tabular}}
\end{table}

In Table~\ref{tab:devices-overview}, we provide an overview of the devices and their locations involved in the two rounds of our experiments.
For each location, we also indicate whether we use WiFi~(W), or cellular~(4G/4G+/5G) connections, or both for measurements at the respective location.
Additionally, Table~\ref{tab:location-distances} lists distances between locations for all three setups.

\begin{table}[tb]
    \centering
    \setlength{\tabcolsep}{1.5pt}
    \caption{Distances [km] between device locations.}
    \label{tab:location-distances}
    \resizebox{\columnwidth}{!}{
    \begin{tabular}{@{}lrrrrc|@{\hskip 0.2cm}lrrrc|@{\hskip 0.2cm}lrrrr@{}}
    \toprule
    \multicolumn{5}{l}{\emph{Round 1}} && \multicolumn{4}{l}{\emph{Round 2 (UAE)}} && \multicolumn{5}{l}{\emph{Round 2 (Germany)}}\\
    \midrule
    & \rotatebox{70}{DE-B} & \rotatebox{70}{NL-A} &\rotatebox{70}{GR-A}& \rotatebox{70}{AE-A}&&& \rotatebox{70}{AE-B} & \rotatebox{70}{AE-C} & \rotatebox{70}{AE-D} &&& \rotatebox{70}{DE-B} & \rotatebox{70}{DE-C} & \rotatebox{70}{DE-D} & \rotatebox{70}{DE-E}\\
    \midrule
    DE-A & 1.5 & 98.7 & 1972.9 & 4981.0 && AE-A & 7.8 & 0.4 & 19.3 && DE-A & 1.5 & 14.4 & 3.4 & 5.4\\
    DE-B & & 97.5 & 1974.4 & 4982.2 && AE-B & & 8.1 & 24.9 && DE-B & & 13.5 & 2.3 & 4.0\\
    NL-A & & & 2065.8 & 5079.5 && AE-C & & & 18.9 && DE-C & & & 11.2 & 10.3\\
    GR-A & & & & 3263.3 &&&&&&& DE-D &&&& 2.3\\
    \bottomrule
    \end{tabular}}
    
\end{table}

\subsection{Measurement Procedure}
\label{sec:msg-measurement-proc}

We measure the time it takes for a message from a sender device to be delivered to the messenger server and to the recipient. 
To this end, we capture an active smartphone's network traffic directly on the device using the \emph{tPacketCapture} app. 
The phone is connected to a computer via USB and a Python script controlling the phone via \emph{Android Debug Bridge (ADB)} automatically schedules the processes of sending messages and capturing network traffic.
The script uses system commands to open and close the packet capture and messaging apps, and interacts with the UI to navigate within the apps, \ie, simulates human touch input to select contacts or type messages.

In a single experiment iteration, the phone subsequently sends a series of five messages to all receivers, with each messenger that is running on the sender and on the receiver device.
The texts of the messages remain the same throughout the whole experiments. 
The first four messages are short and only consist of a single word%\footnote{Marmorkuchen, Sahnetorte, Apfelstreusel, Spekulatius} 
each, while the last message is a whole text paragraph. 
We send the first four messages at an interval of 10~seconds to allow for the confirmations to arrive before sending the next message, while we increase the waiting time before the last messages to 20~seconds in order to accommodate the longer time it takes to type the long text, thus facilitating the analysis of the packet captures.
The measurement procedure is complete when all iterations have terminated successfully for all recipients and their corresponding messaging applications. 
Algorithm~\ref{alg:meas_procedure} shows our measurement procedure.

\begin{algorithm}[h!]
\SetKwInOut{Input}{input}\SetKwInOut{Output}{output}
\SetKw{KwIn}{in}\SetKwFor{For}{for}{\string:}{}
\SetKw{KwTo}{to}\SetKwFor{For}{for}{\string:}{}
\SetKwFunction{startpcap}{start\_pcap}
\SetKwFunction{startapp}{start\_app}
\SetKwFunction{openchat}{open\_chat}
\SetKwFunction{send}{send}
\SetKwFunction{sleep}{sleep}
\SetKwFunction{closeapp}{close\_app}
\SetKwFunction{stoppcap}{stop\_pcap}

\Input{A list of $messengers$ which are supported applications of the receivers}
\Input{A list of $receivers$ according to the contact list}
\Input{A list of $words$ which are sent to the receivers consecutively}
\Output{$void$ function}

$sleep\_time$ = 10\;
$num\_of\_messages$ = 5\;
\For{$receiver$ \KwIn $receivers$}{
  \For{$messenger$ \KwIn $messengers$}{
     \startpcap()\;
     \startapp($messenger$)\;
     \openchat($receiver$)\;
     \For{$i\leftarrow 0$ \KwTo $num\_of\_messages$ - 2}{
        \send(words[i])\;
        \sleep($sleep\_time$)\;
     }
     \sleep($sleep\_time$)\;
     \tcc{Send the long text}
     \send(words[$num\_of\_messages$ - 1])\;
     \closeapp($messenger$)\;
     \stoppcap()\;
  }
}
\BlankLine
%\Return $0$;
\caption{Texting Thumb}
\label{alg:meas_procedure}
\end{algorithm}
\DecMargin{1.5em}

We repeat this procedure over a period of several weeks in July and August 2021 for Round~1 and March to April 2022 for Round~2.
Whereas the physical locations of receiving devices remain unchanged throughout the Round~1 measurements, we collect data for at least one week for each location a receiving device was placed at in Round~2.
In total, we use more than \num{240000} messages sent during the two rounds of experiments for evaluation.

% R1: 65605 messages
% R2: 176210 messages

\section{Descriptive Dataset Analysis}
\label{sec:evaluation}

Using the setup described in Section~\ref{sec:experiments}, we collected our dataset and use it in the further investigations.\footnote{The raw data contain private location data we prefer not to share publicly.} 

\subsection{Data Processing}
\label{sec:eval-data-proc}
For each measurement iteration, we evaluate the recorded packet captures to determine the elapsed time between a message sent by the sender and the notifications (by the server and receiver) that return to the sender.

Since the messengers we consider use multiple layers of encryption (\ie, end-to-end encryption between the communication partners and TLS-encryption for connections between clients and servers on the transport layer), we are not able to access the contents of the communication. 
Yet to analyze the communication flow and identify the messages and confirmations, we develop heuristics from sample captures.
We analyze characteristics of the network traffic such as packet sizes, their order and flow direction, which is a common technique, \eg, for traffic analysis~\cite{cheng98:traffic-analysis-ssl,sun02:statistical-identification-encrypted}.

Within our analysis, we only consider traffic between the sender device and IP addresses within the IP address range(s) of the respective messaging service (\cf~Section~\ref{sec:messenger-infrastructures}).
We are interested in sequences of packets of the form as illustrated in the information flow overview in Figure~\ref{fig:info_flow}. 
The message sent by the sender usually consists of one or more outgoing TCP packets whose destination is one of the messenger servers.
After a message has been sent, there is one incoming TCP packet containing the server notification, coming from the messenger server.
Finally, once the receiver has confirmed that they have retrieved the message, there is another incoming TCP packet containing the delivery notification. 
From the sender's perspective, this packet is also coming from the messenger server.
These observations are based on a first manual visual inspection of a small set of packet capture files. 

Taking into account the aforementioned network traffic structure, we conduct our detailed packet capture analysis in two steps:
\begin{compactenum}[(1)]
\item Identifying typical packet sizes of server and receiver notifications.
\item Matching sequences of TCP packets to determine round-trip times between sending a message and receiving the notifications.
\end{compactenum}

\subsubsection{Identifying Packet Sizes of Notifications}
In the first step, we use a subset of $n=1000$ randomly selected packet capture files and analyze the packet sizes of the two types of incoming packets (\ie, the notifications from server and receiver).
To make sure that we only consider packets that contain these notifications, we limit our first analysis to sequences of packets that appear right after one another and right after the message has been sent.

We then analyze the lengths of the two inbound packets in all matched packet sequences across all packet capture files to identify the lengths of the packets containing the two types of notifications.
We evaluate the frequencies of packet lengths, conducting an additional round of manual plausibility checks within the traces.
The results are listed in Table~\ref{tab:notification-packet-lengths}.
Most notably, the length of the packet containing the notification that a message has been delivered to its receiver in Threema is uniformly distributed between \num{158} and \num{390} bytes. In contrast, the other notifications have smaller variations in packet length: Signal's notifications range from \num{773} to \num{828}, and WhatsApp's from \num{61} to \num{62}.

\begin{table}[t]
\begin{center}
\caption{TCP packet lengths of notifications.~\label{tab:notification-packet-lengths}}
\begin{tabular}{lcc}
\toprule
Messenger & Bytes (Server) & Bytes (Receiver)\\
\midrule
Signal & \num{123}--\num{124} & \num{773}--\num{828}\\
Threema & \num{38} & \num{158}--\num{390} \\
WhatsApp & \num{68}--\num{69} & \num{61}--\num{62} \\
\bottomrule
\end{tabular}
\end{center}
\end{table}

\definecolor{redline}{RGB}{184,84,80}
\definecolor{redfill}{RGB}{248,206,204}

\pgfdeclarelayer{background}
\pgfsetlayers{background,main}
\begin{figure}[t]
\centering
\begin{adjustbox}{width=.9\columnwidth}
\begin{tikzpicture}[
    sectionnode/.style={text width=3cm, text centered,font=\sffamily\small},
	squarednode/.style={rectangle, draw=gray!60, fill=gray!5, very thick, minimum size=8mm, text width=8em, text centered, font=\sffamily\small},
	roundednode/.style={rectangle, draw=gray!60, fill=gray!5, very thick, rounded corners, minimum size=8mm, font=\sffamily, text width=10em, text centered},
	attacknodered/.style={rectangle, draw=redline, fill=redfill, very thick,  minimum size=8mm, font=\sffamily\footnotesize, text width=8em, text centered},
	textnode/.style={minimum size=8mm, text width=3em, text centered},
	font=\sffamily\large,
	packet/.style={rectangle, draw=gray!20, fill=gray!2, very thick, minimum size=4mm, text=gray!50, text width=24em, text centered, anchor=west, font=\sffamily\small},
	packetmatch/.style={rectangle, draw=gray!60, fill=gray!5, very thick, minimum size=4mm, text width=24em, text centered, anchor=west, font=\sffamily\small},
    ]
	%LEFT BLOCK
	\node[packet] (p0) at (0,0) {\parbox{10cm}{\emph{idx=207, t=53.9259, dir=outbound, len=536}}};
	\node[packet] (p1) at (0,-0.6) {\parbox{10cm}{\emph{idx=208, t=53.9261, dir=inbound, len=42}}};
	\node[packetmatch] (p2) at (0,-1.2) {\parbox{10cm}{\emph{idx=209, t=53.9263, dir=outbound, len=97}}};
	\node[packet] (p1) at (0,-1.8) {\parbox{10cm}{\emph{idx=210, t=53.9264, dir=inbound, len=42}}};
	\node[packetmatch] (p1) at (0,-2.4) {\parbox{10cm}{\emph{idx=211, t=54.0722, dir=inbound, \textbf{len=123}}}};
	\node[packet] (p1) at (0,-3.0) {\parbox{10cm}{\emph{idx=212, t=54.1225, dir=outbound, len=42}}};
	\node[packetmatch] (p1) at (0,-3.6) {\parbox{10cm}{\emph{idx=213, t=55.0154, dir=inbound, \textbf{len=776}}}};
	\node[packet] (p1) at (0,-4.2) {\parbox{10cm}{\emph{idx=214, t=55.0656, dir=outbound, len=56}}};

	\node[textnode] (m) at (8,-1.2) {$m$};
	\node[textnode] (m) at (8,-2.4) {$n_{1}$};
	\node[textnode] (m) at (8,-3.6) {$n_{2}$};
	%\node[packet] (p3) at (0,-1.8) {\parbox{10cm}{\emph{idx=3, t=54.0555, dir=inbound, len=42}}};
	%\node[packet] (p4) at (0,-2.4) {\parbox{10cm}{\emph{idx=4, t=54.0555, dir=inbound, len=42}}};
	%\node[packet] (p5) at (0,-3.0) {\parbox{10cm}{\emph{idx=5, t=54.0555, dir=inbound, len=42}}};
	%\node[packet] (p6) at (0,-3.6) {\parbox{10cm}{\emph{idx=6, t=54.0555, dir=inbound, len=42}}};
	%\node[packet] (p7) at (0,-4.2) {\parbox{10cm}{\emph{idx=7, t=54.0555, dir=inbound, len=42}}};

\end{tikzpicture}
\end{adjustbox}
\caption{Excerpt from an example packet capture with the three identified packets of interest highlighted.}
\label{fig:packet_matching}
\end{figure}

\subsubsection{Matching Packet Sequences to Determine RTTs}

In the second step, we systematically analyze all packet captures we have collected during the two rounds of measurements. 
Since we now know the sizes of packets we are interested in, we omit the requirement of packets to appear right after one another in the correct order.
This helps us to also identify messages whose delivery notification is delayed, or when the traffic patterns we are interested in interferes with other packets exchanged between the client application and the messenger server.
We first identify the two inbound packets (\ie, the two notifications $n_{1}$ and $n_{2}$) based on their size and match them with the latest outbound packet (\ie, the message~$m$) sent before those two packets arrived.
An example is illustrated in Figure~\ref{fig:packet_matching}.
We use the timestamps of the three packets (\ie, $t(m)$ for message~$m$) to determine the notification round-trip times (RTT) between (S)ender and (M)essenger Server, and (S)ender and (R)eceiver:
\begin{equation}
\begin{split}
RTT_{S,M} = t(n_{1}) - t(m)
\\
RTT_{S,R} = t(n_{2}) - t(m).
\end{split}
\end{equation}

\noindent
Finally, we calculate the hypothetical RTT between (M)essenger Server and (R)eceiver:
\begin{equation}	
RTT_{M,R} = RTT_{S,R} - RTT_{S,M}.
\end{equation}

\subsubsection*{Additional Notes on Signal in the UAE}

In the Signal data collected in Round 2 in the UAE, we observed different traffic characteristics. 
In particular, there is only one specific packet returned from the server~--~presumably containing both confirmations from server and receiver. 
Thus, we cannot determine the difference between the two but only consider $RTT_{S,R}$ for our analysis.

\subsection{Delivery Notification Timings}
\label{sec:eval-timings}

We now present a first view into our delivery notification timing dataset.
We start by analyzing the measured times in relation to the traveled distance, and later continue with distributions of timings to different receivers.

\subsubsection{Timings and Distances}
We are first interested in the relation between the timings we observe and the traveled distances between sender, messenger server, and receiver.
To this end, we analyze what messenger servers have been picked on the sender's side and leverage the findings from our messenger infrastructure analysis (\cf~Section~\ref{sec:messenger-infrastructures}) to determine the distances from the server to sender ($dist_{GCD}(S,M)$) and receiver ($dist_{GCD}(M,R)$), respectively.
We emphasize that the receiving device might be connected to a different server (location) than the sender~--~however, from the attacker's position (\ie, the sender), this information cannot be further resolved.
We can then analyze the relation between timings and distances for the two segments.

\begin{figure}[t]
\centering
         \centering
         \includegraphics[width=\columnwidth]{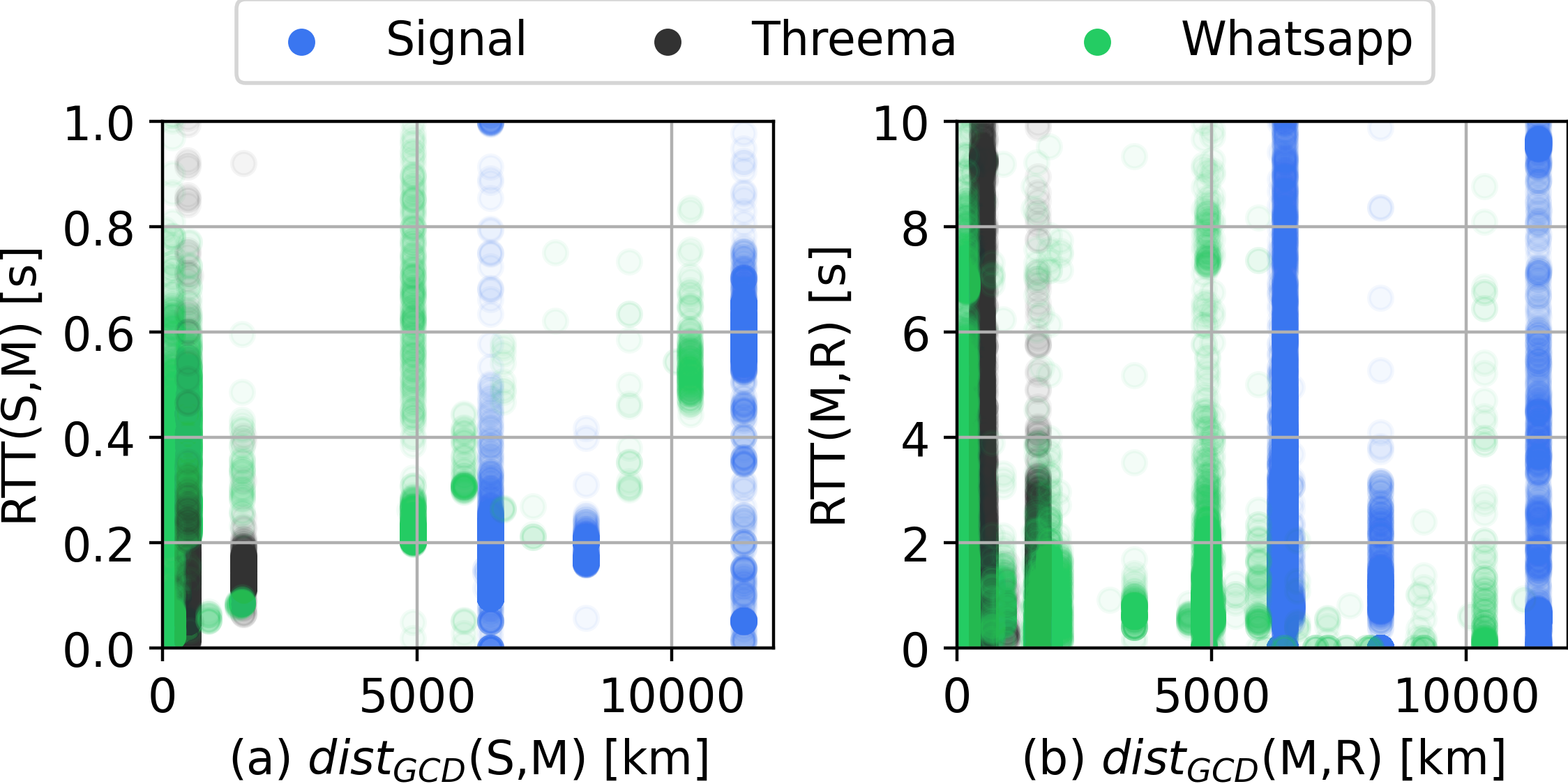}
         \caption{Round trip time distributions of distance splits for (a) sender to server and (b) server to receiver~--~(each with \num{10000} randomly sampled timings per measurement round)~\label{fig:all_distances}}
\end{figure}

\begin{figure}[t]
\centering

\begin{subfigure}{\columnwidth}
\includegraphics[width=\columnwidth]{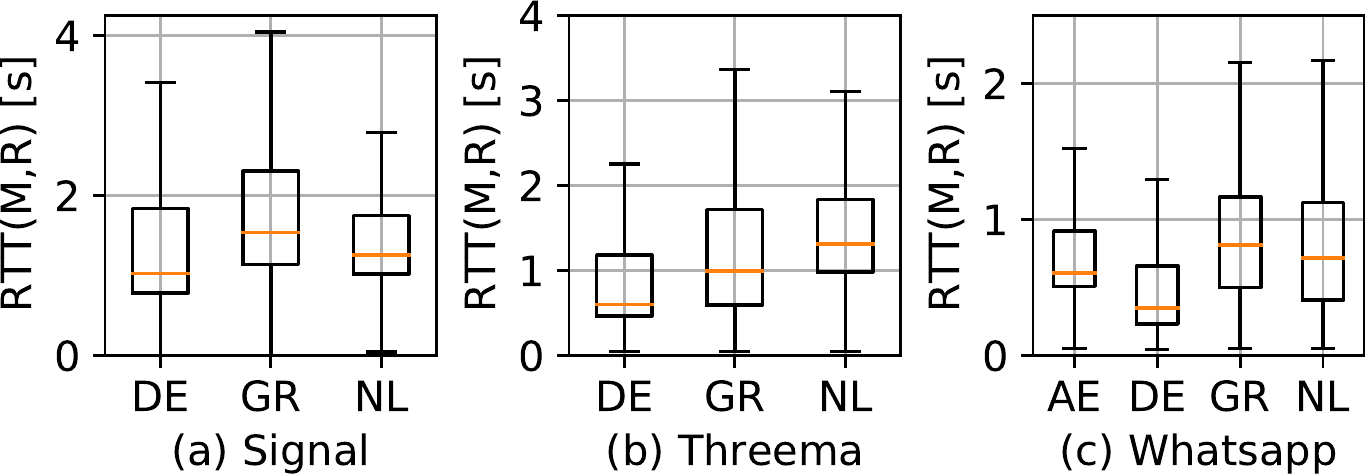}
\caption{Messages sent from device \emph{DE-11} to receivers in different countries. Y-axes have different ranges since we only intend to highlight differences within each messenger.~\label{fig:msg_timings-countries}}
\end{subfigure}

\vspace{0.2cm}

\begin{subfigure}{\columnwidth}
\includegraphics[width=\columnwidth]{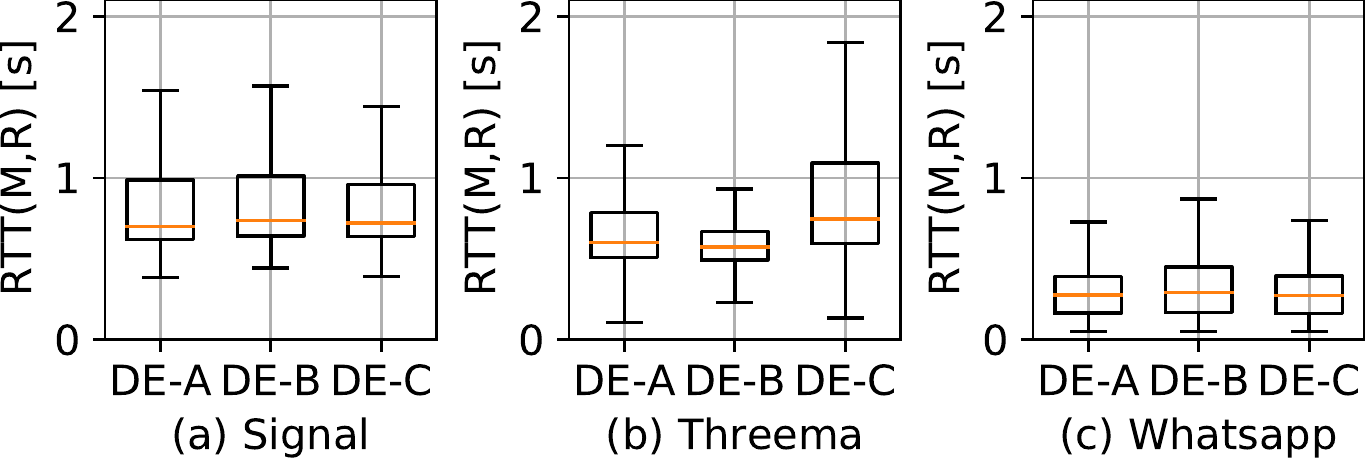}
\caption{Messages sent to device \emph{DE-22} separated by the device's location.} 
\label{fig:box_3wloc1_de-p8}
\end{subfigure}

\vspace{0.2cm}

\begin{subfigure}{\columnwidth}
\includegraphics[width=\columnwidth]{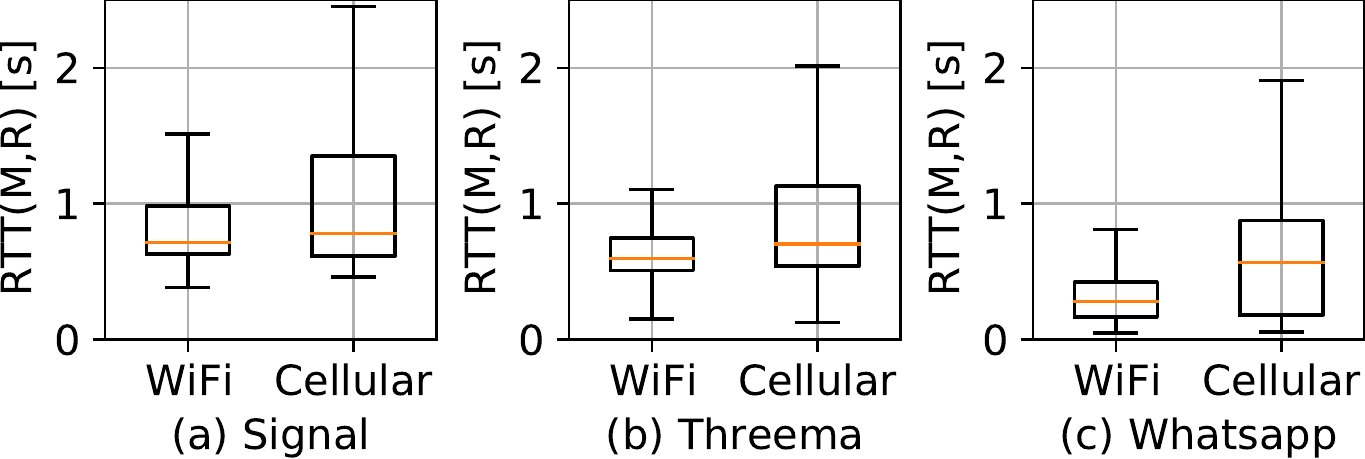}
\caption{Messages sent to device \emph{DE-22} separated by its network connection.} 
\label{fig:box_network_de-p8}
\end{subfigure}

\caption{Empirical distributions of $RTT_{M,R}$ [s] for messages sent at different stages of our experiments.}
\label{fig:dataset-boxplots}

\end{figure}

In Figure~\ref{fig:all_distances}a, we see a slight tendency for minimum timings to increase for longer distances between sender and server (for Threema and Whatsapp), even though timings are largely scattered for similar distances.
In Figure~\ref{fig:all_distances}b, there is, again, a comparably small set of distances between servers and receivers, and timings being scattered a lot without clear trends.
Since our experiments only cover a small set of distances between devices, and only consider Great Circle Distances (GCD) between entities, without taking into account the actual routing through the Internet topology, our dataset does not allow to develop a generalized model to put timings in direct relation to the traveled distances.
To reduce the noise introduced into our data at this stage, we continue with focusing on the timings between messenger server and receiver, \ie, we use $RTT_{M,R}$ in subsequent analyses.

\subsubsection{Differences between Receiver Characteristics}
In the next step, we analyze to what extent timings we collected comprise differences between receivers, or their characteristics, respectively.

We first compare the measured $RTT_{M,R}$ between receivers the different countries involved in the first round of experiments.
Figure~\ref{fig:msg_timings-countries} illustrates distributions of these timings of messages sent from device \emph{DE-11} to receivers in different countries for each messenger. 
For all messengers, we observe that timings to Germany are shorter (lower medians) and tighter distributed (smaller boxes). 
Shorter timings for Germany are the result we expect in this case, since all messages have also been sent from a device in Germany.
Whereas the differences between the medians are smaller for the other countries, distributions have different widths (heights of boxes) or are differently shifted (position of boxes).

While differences between the distributions of notification timings to receivers in different countries can be easily identified, we also analyze if such differences also exist at smaller scale.
Moreover, we cannot entirely exclude that these differences are partially grounded in the devices itself, since in the first round of measurements, each country location corresponds to a different device.
In this regard, we now compare notification timings of messages sent to device DE-22 at its different locations in Germany during the second round of measurements.
Figure~\ref{fig:box_3wloc1_de-p8} shows the distributions of timings to the three locations.
Differences appear to be much smaller than those on the per-country level, we can only observe small variations in, \eg, medians or ranges of timing distributions, indicated by ranges and shapes of boxes.

In the last step, we also compare notification timings sent to the same device depending on its network connection.
In this case, differences appear to be larger again, with distributions of timings of messages received over cellular data showing a higher variance (larger box) and being slightly slower, indicated by a higher median (\cf~Figure~\ref{fig:box_network_de-p8}).

\section{Delivery Notification Timing Classification}
\label{sec:classification}

Classifying the timing measurements collected in the experiments can help to determine certain characteristics of the receiver of a message, such as their location.
We demonstrate at what scale it is feasible to classify different targets based on delivery notification timing measurements and to distinguish these characteristics from each other.

\subsection{Classification Tasks}
\label{sec:classification-tasks}
To evaluate and demonstrate at what scale the classification of receivers and their characteristics works, we specify a set of classification tasks at different granularity levels as follows:

\begin{compactenum}[(1)]
\item \emph{Country}: We distinguish our measurements by the country a receiving device is located in (out of the set of countries we have measurements for).
\item \emph{Within Country}: We only distinguish whether or not a receiving device is located within a specific country.
\item \emph{City Location}: We distinguish timings to different locations within the same city. We repeat this classification task for devices individually and conjointly.
\item \emph{Network Connection}: We distinguish whether a device is using a WiFi or a cellular Internet connection.

\end{compactenum}

According to the designs of our measurement setups (\cf~Section~\ref{sec:measurement-setup}), we use data from the first round of measurements for classification tasks~(1) and~(2), whereas classification tasks~(3) and~(4) are based on data from the second round of measurements.

\subsection{Classification Setup and Parameter Tuning}
We use \emph{sequences} of delivery notification timings for classification. 
A sequence is a set of notification timings derived from up to five subsequently sent messages (\cf~Section~\ref{sec:msg-measurement-proc}).
We repeat the classification with different sequence lengths, starting with $n=1$, \ie, a single notification round-trip time from a single message.

For each classification task, we analyze the measurement data for each sender device and for each messenger independently. 
We randomly sample \emph{k} notification timing sequences from each class, whereas \emph{k} is the number of timing sequences of the class with the lowest number of sequences. This way, we reach an evenly weighted set of samples per class.

We use convolutional neural networks (\emph{CNN}) as classifiers, train them with sequences of delivery notification timings from different classes and then measure their accuracy in predicting newly observed timing sequences.
This selection is grounded in our own preliminary parameter tuning evaluation and builds upon findings by Rimmer~\etal~\cite{rimmer18:automated-website-fingerprinting}, who extensively evaluate the performance of different types of neural networks for a similar network traffic analysis task (Website Fingerprinting) and report CNNs to perform best when compared to Long Short-Term Memory (LSTM) networks and Stacked Denoising Autoencoder (SDAE) networks.
Before we start the actual classifications, we repeatedly run the first classification task with varying parameters to find the optimal classification setup for each of the three types of neural networks, \ie, CNN, LSTM, and SDAE and compare the results.
Details about the parameter tuning configurations can be found in Appendix~\ref{sec:appendix-nn-tuning}.

\subsection{Classification Procedure and Evaluation Metrics}
For each classification task, we randomly split the respective data into five portions and use all but one of these portions as training set for the neural network.
The remaining portion serves as test set from which all samples are to be classified.
For each sample in the test set, the neural network output comprises a softmax result, \ie, assigning each candidate class a probability that the classified sample belongs to this class.
Based on the softmax output, we assign each sample the class with the maximum probability, considering this as the classification decision.
To avoid model over-fitting, we repeat this procedure until each of the five data portions has served as test set and merge the five classification results, effectively implementing 5-fold cross-validation.

The performance of the classification is determined by the numbers of classifications that identify the correct class (\emph{precision}) and the number of samples in each class that are correctly classified (\emph{recall}). 
In our evaluation, we focus on precision, \ie, we are interested in the fraction of samples per class that can be correctly identified and how the classifications are distributed for all samples of a particular class.
We also analyze changes in classification performance when we vary the sequence length.

We report these detailed results for the first classification task (\ie, distinguishing receiver countries) to provide detailed insights into our evaluation and how it works.
For subsequently presented classification tasks, we report overall accuracy results for a large number of different classifications using the maximum delivery notification sequence length (\ie, 5 messages).
We do so to provide a broad overview of the varying effectiveness of leveraging the timing side channel in different scenarios. 
%\emph{For interested readers, we provide detailed results of all instances of all classification tasks in Appendix~\ref{sec:appendix-detailed}.}

Finally, we also analyze the convergence of the classification accuracy depending on the sample size, \ie, we repeat a selected set of classifications multiple times with increasing numbers of samples per class, and measure the resulting performance.

\begin{figure}[tb]
\centering
\includegraphics[width=\columnwidth]{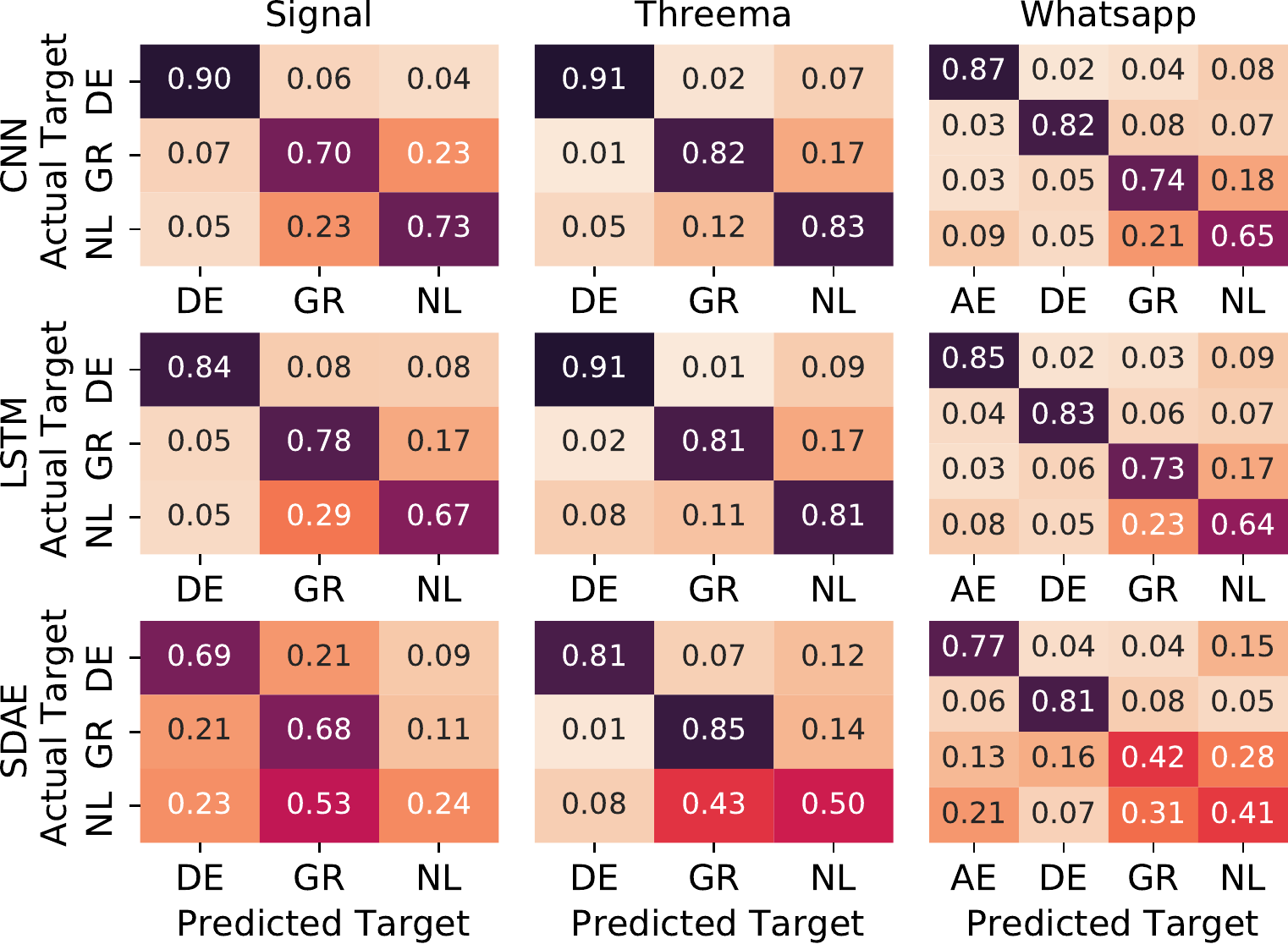}
\caption{Detailed classification results for the receiver \emph{country} based on measurements from sender \emph{DE-11} with three different neural network types. For each classification, numbers report \emph{precision} values, \ie, the fractions of predicted classes (x-axis) given the actual class (y-axis).}
\label{fig:msg_timings-confusion-countries}
\end{figure}

\subsection{Receiver Classification by Country}
\label{sec:target-classification-country}

In the first step, we present the results of the receiver \emph{country} classification for one sender device in Germany (\emph{DE-11}).
For WhatsApp, the receiver can be one of four countries (\emph{AE}, \emph{DE}, \emph{GR},  \emph{NL}). For the two other messengers, we cannot present data for \emph{AE} due to the messenger not being available at all in the country (Threema) or too little successful delivery notification measurements (Signal). To this end, we are restricted to the three remaining countries for Signal and Threema.

\begin{figure}[tb]
\centering
\includegraphics[width=\columnwidth]{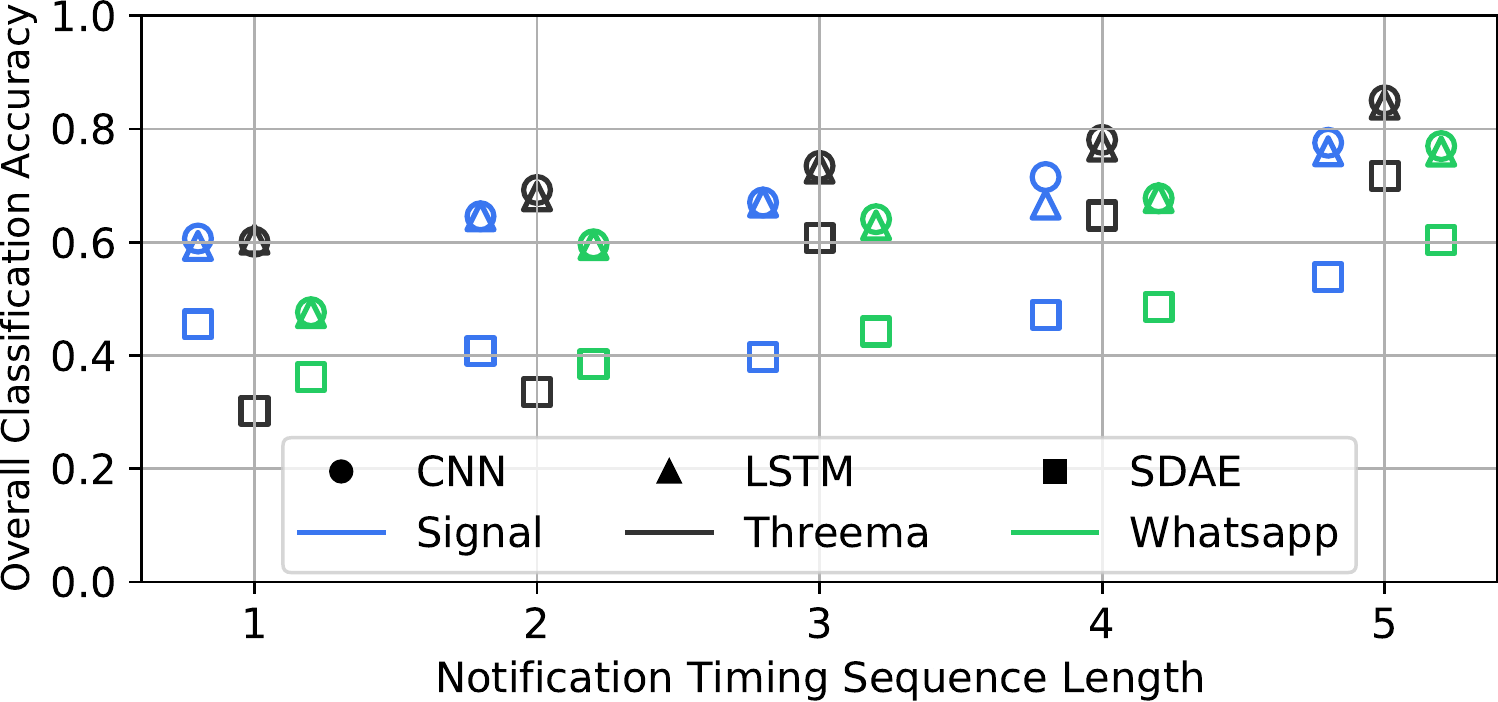}
\caption{Overall classification accuracy (y-axis) for the receiver classification per country, depending on delivery notification timing sequence length (x-axis) and NN type (icon shape)}.
\label{fig:msg_timings-accuracy-countries}
\end{figure}

Detailed results are presented in confusion matrices in Figure~\ref{fig:msg_timings-confusion-countries}, separated by messenger (columns) and neural network type\ (rows).
The numbers indicate the fractions of predicted classes for each actual class (\emph{precision} values).
A darker principal diagonal in each matrix indicates higher accuracy since numbers on this axis refer to correct predictions.
Figure~\ref{fig:msg_timings-accuracy-countries} illustrates corresponding overall accuracy for this classification tasks for all three messengers depending on the length of the notification timing sequence.

For Signal (left column matrices), the receivers located in Germany can be distinguished from receivers in the two other countries quite well. 
We observe false classifications mostly between devices in GR and NL.
This result is not surprising since timing distributions for GR and NL largely overlap, whereas timings of messages to receivers in DE are lower (\cf~Figure~\ref{fig:msg_timings-countries}).
The overall classification accuracy rises from \SI{60}{\percent} for a single timing per sample to \SI{79}{\percent} for \num{5} timings per sample~(\cf~Figure~\ref{fig:msg_timings-accuracy-countries}) in the case of a CNN classification.
For Threema, there is a quite similar outcome. 
Again, classification works best for receivers in \emph{DE} with most false classifications between \emph{GR} and \emph{NL}.
For longer timing sequences, Threema reaches a better overall accuracy of \SI{86}{\percent} for \num{5} timings per sample, compared to \SI{60}{\percent} for single-time samples.
In the case of WhatsApp, receivers in \emph{DE} and \emph{AE} can be distinguished best from the others and performance increases for longer timing sequences.
The overall accuracy is a bit lower for the other two messengers (\ie, \SI{47}{\percent} to \SI{77}{\percent}).

Regarding the classifier type, CNN and LSTM perform with similar quality with CNN reaching slightly higher performances in most cases.
SDAE results are noticeably worse.
Therefore, and taking into account previous findings~\cite{nasr18:deepcorr-strong-flow,rimmer18:automated-website-fingerprinting}, we continue with CNN throughout the remaining evaluations.

\begin{table}[t]
\caption{Detailed precision results for the classification of receiver countries (CNN-based classification). 
\label{tab:classification-accuracy-details}}

%\katha{how many measurements for each cell? $\rightarrow$ define a confidence score or metric that explains how convincing the results of a row/cell are}.
%\centering

\resizebox{1.\columnwidth}{!}{
\begin{tabular}{@{}lgggrrrgg}
\toprule
    Sender  & \multicolumn{3}{h}{\textbf{DE-11}} & \multicolumn{3}{c}{\textbf{DE-12}} & \multicolumn{2}{h}{\textbf{GR-11}} \\
    \midrule
    Messenger & \multicolumn{1}{h}{SIG} & \multicolumn{1}{h}{THR} & \multicolumn{1}{h}{WA} & \multicolumn{1}{c}{SIG} & \multicolumn{1}{c}{THR} & \multicolumn{1}{c}{WA} & \multicolumn{1}{h}{THR} & \multicolumn{1}{h}{WA} \\
    \midrule
% DO NOT EDIT THIS TABLE MANUALLY!
\multicolumn{8}{l}{\emph{Classification Task: Country}}\\
AE & -- & -- & 84\,\% & -- & -- & 94\,\% & -- & 95\,\%\\
DE & 90\,\% & 94\,\% & 81\,\% & 73\,\% & 70\,\% & 77\,\% & 71\,\% & 89\,\%\\
%DK & -- & -- & -- & -- & -- & -- & -- & -- & --\\
GR & 77\,\% & 84\,\% & 79\,\% & 53\,\% & 68\,\% & 64\,\% & -- & --\\
NL & 70\,\% & 80\,\% & 63\,\% & 61\,\% & 68\,\% & 53\,\% & 66\,\% & 88\,\%\\
\addlinespace[0.1cm]
\emph{Samples/Class} &\multicolumn{1}{h}{$177$} &\multicolumn{1}{h}{$527$} &\multicolumn{1}{h}{$825$} &\multicolumn{1}{c}{$66$} &\multicolumn{1}{c}{$60$} &\multicolumn{1}{c}{$135$} &\multicolumn{1}{h}{$187$} &\multicolumn{1}{h}{$168$}\\
\emph{Overall Acc.} &79\,\% &86\,\% &77\,\% &62\,\% &69\,\% &72\,\% &68\,\% &90\,\%\\
\midrule
% DO NOT EDIT THIS TABLE MANUALLY!
\multicolumn{8}{l}{\emph{Classification Task: Within Germany}}\\
DE & 92\,\% & 91\,\% & 90\,\% & 86\,\% & 84\,\% & 92\,\% & 90\,\% & 90\,\%\\
NOT-DE & 91\,\% & 94\,\% & 92\,\% & 78\,\% & 85\,\% & 88\,\% & 51\,\% & 94\,\%\\
\addlinespace[0.1cm]
\emph{Samples/Class} &\multicolumn{1}{h}{$559$} &\multicolumn{1}{h}{$1135$} &\multicolumn{1}{h}{$1888$} &\multicolumn{1}{c}{$250$} &\multicolumn{1}{c}{$180$} &\multicolumn{1}{c}{$605$} &\multicolumn{1}{h}{$187$} &\multicolumn{1}{h}{$349$}\\
\emph{Overall Acc.} & 91\,\%&92\,\%& 91\,\%&82\,\%&85\,\%&90\,\%&70\,\%&92\,\%\\

\bottomrule
\end{tabular}
}
\end{table}

Table~\ref{tab:classification-accuracy-details} lists precision results per class for the country classification for messages sent with all three messengers from three sender devices (\emph{DE-11}, \emph{DE-12}, and \emph{GR-11}).
We also report sample sizes of notification timing sequences there.
All results listed in the table refer to the maximum notification timing sequence length, \ie, timings of \emph{five} subsequently sent messages.
The results in the top left block of the table correspond to the numbers presented in Figure~\ref{fig:msg_timings-confusion-countries} with each table column corresponding to the principal diagonal axis in the respective confusion matrix.

\subsubsection{Country Subsets}
We repeat the classification of delivery notification timing sequences with the other devices and for every subset of countries in our data set.
The resulting set comprises one more classification of four countries (sender device \emph{DE-12}) and multiple evaluations of all possible pairs and triplets of countries including measurements from all three sender devices.
In this context, we only consider the maximum sequence length, \ie, delivery notification timings of $n=5$ subsequently sent messages.

\begin{figure}[tb]
\centering
\includegraphics[width=\columnwidth]{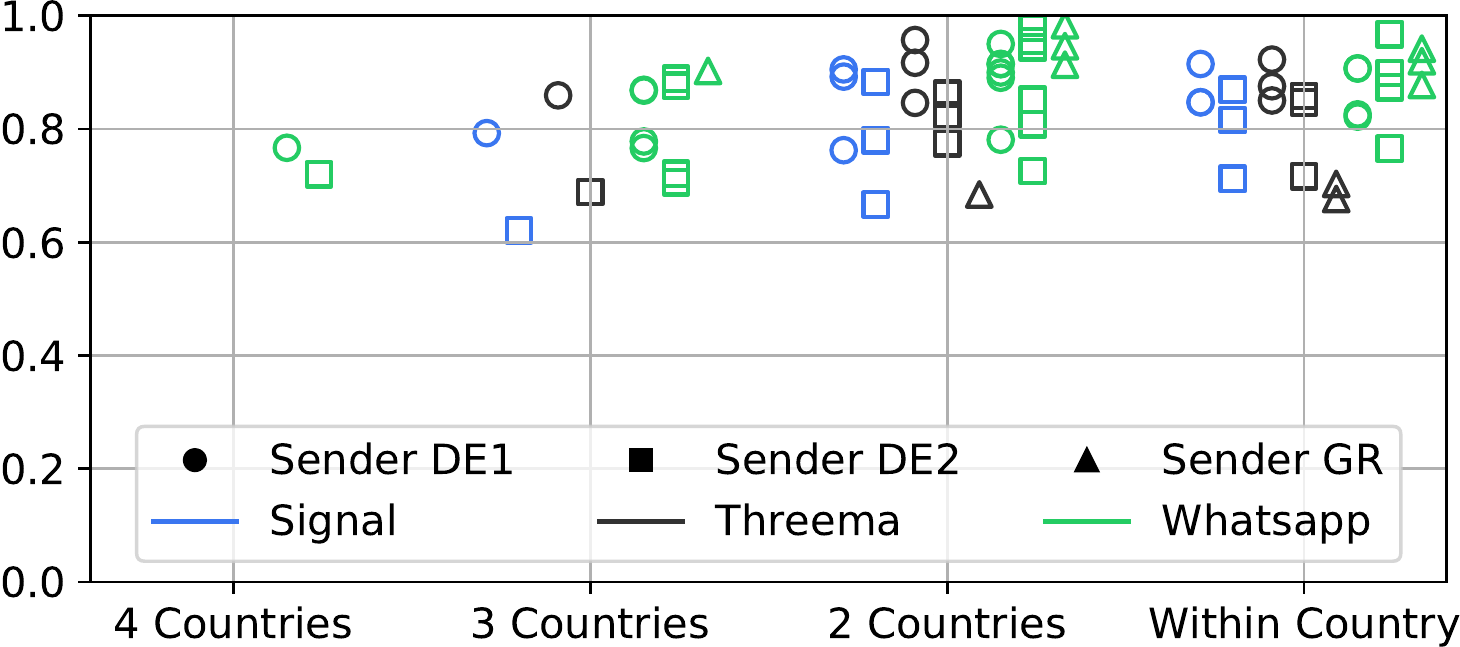}
\caption{Overall accuracy of receiver country classification separately for all possible country subsets for each sender device (icon shape) and messenger (colors).}
\label{fig:msg_timings-accuracy-countries-meta}
\end{figure}

Figure~\ref{fig:msg_timings-accuracy-countries-meta} shows the overall accuracy values of the receiver country classification for all combinations of countries in our data set. 
For smaller target sets, classifications perform better, with overall classification accuracy mostly between \SI{70}{\percent} and \SI{90}{\percent}.
In the case of two countries, some classifications even perform with more than \SI{95}{\percent} accuracy.
Such nearly perfect results can only be achieved when timings can be clearly distinguished, which is mostly the case when the candidate locations are far from each other (one receiving device located in the UAE and the other one in a European country).
However, also for distinguishing notification timings of messages sent to Germany and to the Netherlands (\emph{DE11-2countries1}), we achieve a classification accuracy of more than \SI{90}{\percent} (\SI{92}{\percent} for Threema and \SI{91}{\percent} for Signal and WhatsApp).

\begin{figure*}[tb]
\centering
\includegraphics[width=\textwidth]{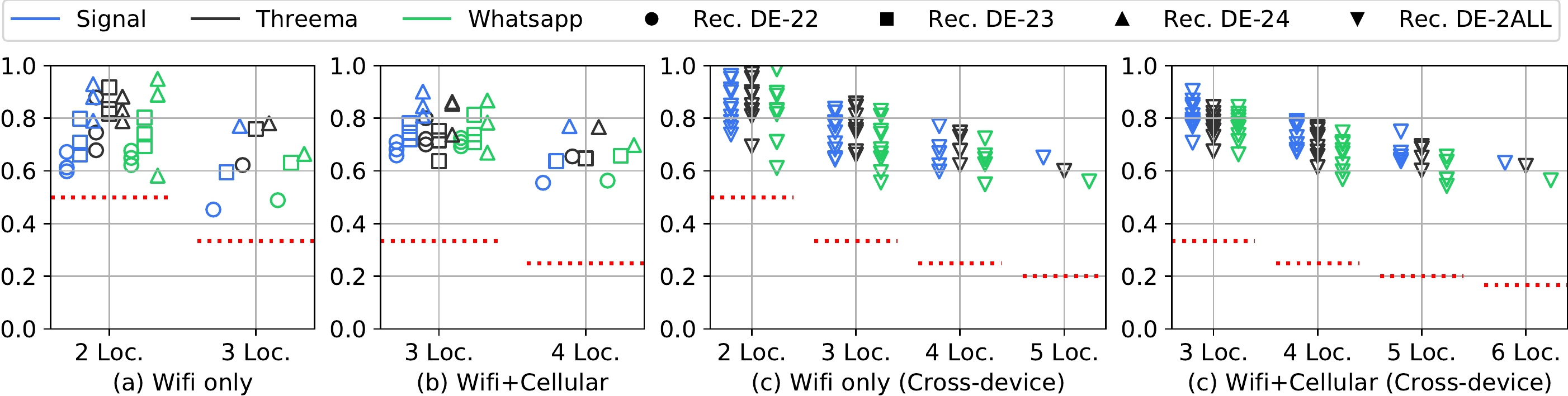}
\caption{Overall accuracy of receiver location classification separately for all possible combinations of locations in Germany. Colors indicate messengers and icon shapes indicate different receiver devices (we refer to Rec. DE-2ALL for the cross-device analysis). Dotted lines indicate the probability of randomly guessing the correct location out of the set of known locations.}
\label{fig:msg_timings-accuracy-r2de-meta}
\end{figure*}

\subsubsection{Within Country}
In the second classification task, we are interested in whether or not a receiver is located in a specific country. 
Different from the previous task, we are not interested in determining the exact location but only in a binary decision about a specific location.
Therefore, we only distinguish notification timing sequences of messages sent to the country we are interested in (\eg, \emph{DE}) from timings to any of the other countries, effectively summarizing timing sequences of all other countries into one class (\eg, \emph{NOT-DE}).
Technically, this type of prediction is similar to the classification of two countries.

Figure~\ref{fig:msg_timings-accuracy-countries-meta} also includes accuracy results for all such classifications, with the majority being very similar to the two-country classification.
As an example, we provide more detailed precision results for the \emph{Within Germany} classification task in Table~\ref{tab:classification-accuracy-details} for all three sender devices.

\subsection{Receiver Locations Within the Same City}
We now present classification results for receivers at different locations within the same city to demonstrate that the timing side channel provided by delivery confirmations also persists at smaller scale.
In this case, the end-to-end distances between sender devices, messenger servers, and receiver devices remain roughly the same across all measurements.
Similar to the per-country classification, we consider all possible combinations of WiFi locations and subsets and evaluate the classification performance for each of them.
Subsequently, we repeat the analysis also including the timing data retrieved from receivers operating on a cellular connection as a separate class.
We repeat these analyses for receiver devices individually and across all devices within the same setup, \ie, the Round 2 measurements in Germany and in the UAE~(\cf~Table~\ref{tab:devices-overview}). 
Whereas cross-device analyses provide first insights towards the generalizability of receiver location classification models (\ie, whether or not the classification requires training for each individual device), the individual analyses ensure that the classification is not biased by timing artifacts introduced by characteristics of the different devices.

\subsubsection{Individual Receivers}
The classification results for the three receiving devices in Germany are illustrated in Figure~\ref{fig:msg_timings-accuracy-r2de-meta}a+b.
The accuracy highly varies between messengers, devices, and the respective combination of locations. Across all combinations of two locations, in each of which the device is connected via WiFi (a), the prediction performance can reach more than \SI{90}{\percent} in some cases, \eg, when distinguishing locations \emph{DE-A} and \emph{DE-B} for the receiver device \emph{DE-24} (\emph{2wloc1-DE-24}).
On the other side of the spectrum, there are also combinations of two locations which cannot be distinguished at all~--~a classification accuracy of around \SI{60}{\percent} is hardly better than randomly guessing one of the two location candidates, \eg, when distinguishing locations \emph{DE-B} and \emph{DE-C} for device \emph{DE-22} (\emph{2wloc5-DE-22}).
For distinguishing three WiFi locations, accuracy is lower with a maximum of \SI{77}{\percent} for Signal, \SI{78}{\percent} for Threema, and \SI{66}{\percent} for WhatsApp (\emph{3wloc3-DE-24}).
However, the chance of randomly guessing one location is also lower in this case (\SI{33}{\percent}).

Identifying the correct location becomes easier when the receiving device operates on a cellular connection in one of them (\cf Figure~\ref{fig:msg_timings-accuracy-r2de-meta}b).
For distinguishing two WiFi locations and one on mobile data, the classification accuracy is mostly between \SI{60}{\percent} and \SI{80}{\percent}.
Such a scenario could, for example, model home and work locations of the device owner, whereas the cellular connection represents any other place in which the phone is not connected to a WiFi network. 

\begin{figure}[t]
\centering
\includegraphics[width=\columnwidth]{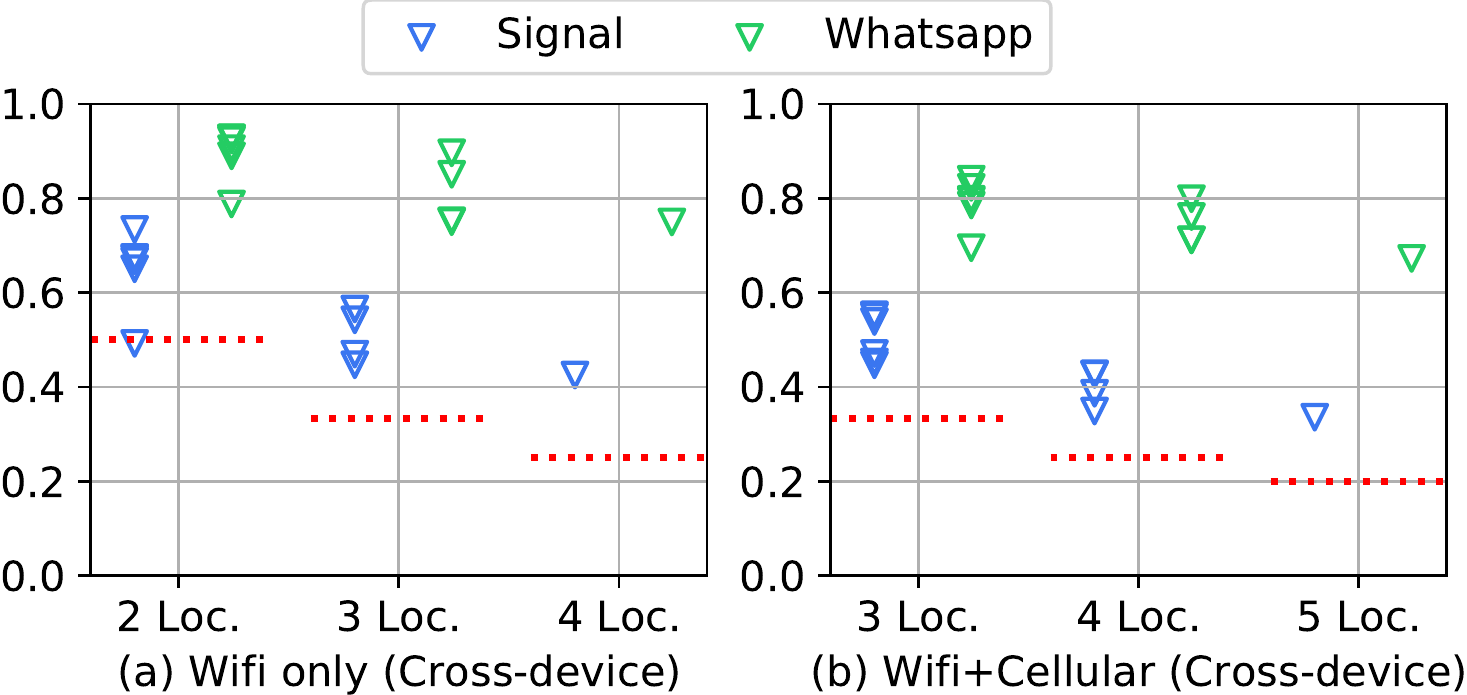}
\caption{Overall accuracy of receiver location classification separately for all possible combinations of locations in the UAE. Colors indicate messengers and dotted lines indicate the probability of randomly guessing the correct location.}
\label{fig:msg_timings-accuracy-r2ae-meta}
\end{figure}

\begin{figure*}[ht]
\centering
\includegraphics[width=\textwidth]{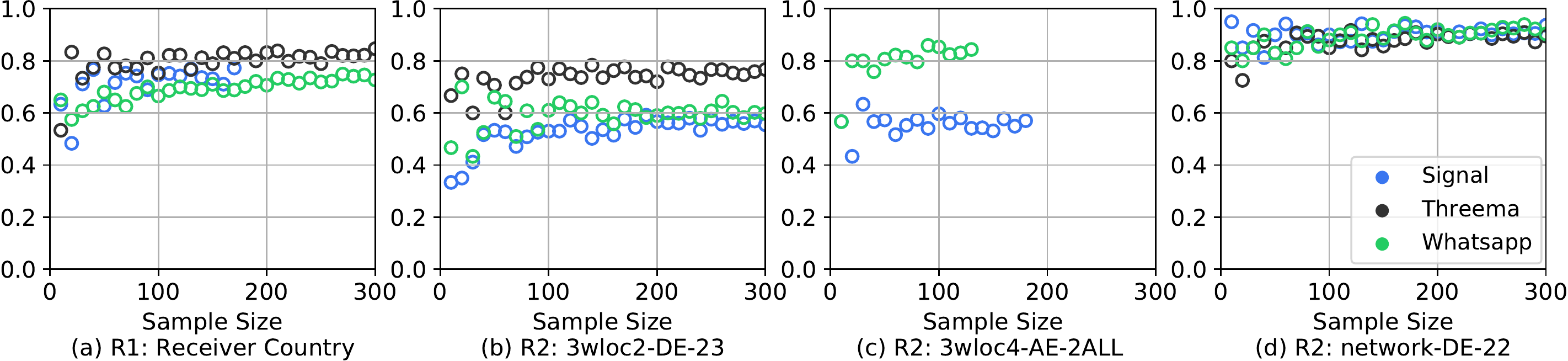}
\caption{Overall accuracy of four different classification tasks, depending on the number of samples per class (x-axis).~\label{fig:model_convergence}}
%\vspace{-0.4cm}
\end{figure*}

\subsubsection{Cross-device Analysis}
When distinguishing locations across different devices (\cf Figure~\ref{fig:msg_timings-accuracy-r2de-meta}c+d), classification performs similar to the case of individual devices, with accuracy increasing slightly.
Such differences might come from individual devices introducing specific timing characteristics into the dataset that facilitate distinguishability of locations. 

For the data collected in the UAE setup, the picture is more diverse.
As the results in Figure~\ref{fig:msg_timings-accuracy-r2ae-meta} show, both two and three WiFi locations can be distinguished with up to more than \SI{90}{\percent} accuracy in WhatsApp, which resembles better performance than comparable classifications in the German setup.
However, for Signal, the classification of locations does not seem to work at all, which we attribute to the different structure of message exchange (and in particular the presence of only one confirmation packet) as described in Section~\ref{sec:eval-data-proc}.

\subsection{Receiver Network Connections}
Since different locations can apparently be better distinguished when the receiving device operates on a mobile data in one of them, we also analyze if we can generally detect whether a phone is connected via WiFi or using a cellular connection.
Being able to distinguish these two cases allows us to determine whether a target is currently in one of their usual locations (\ie, we assume that they are connected to the respective WiFi network there) or not (mobile data).

The results for the evaluation of this classification task are listed in Table~\ref{tab:msg_timings-accuracy-network}.
In the setup in Germany, we can detect the receiver's Internet connection type with high accuracy for all devices for all messengers, both for individual devices and also across different ones.
Classifications reach an overall accuracy of \SI{90}{\percent} or even above, with only one prediction performing worse (Device \emph{DE-23}, Threema).
In the setup in the UAE, predicting the network connection performs on a similar level for WhatsApp. In the case of Signal, results do not seem convincing (\SI{50}{\percent} corresponds to randomly guessing the connection type), which is in line with results of the WiFi location distinguishability.

\begin{table}[b]
    \centering
    \caption{Classification accuracy for receiving devices' network connections (WiFi vs. mobile data)}
    \label{tab:msg_timings-accuracy-network}
    \begin{tabular}{@{}lrrrr@{\hskip 0.8cm}lrr@{}}
    \toprule
    \multicolumn{4}{c}{Germany} && \multicolumn{3}{c}{UAE}\\
    \midrule
    Receiver & \multicolumn{1}{c}{SIG} & \multicolumn{1}{c}{THR} & \multicolumn{1}{c}{WA} && Receiver & \multicolumn{1}{c}{SIG} & \multicolumn{1}{c}{WA}\\
    \midrule
    DE-22 & 92\,\% & 90\,\% & 94\,\% && AE-22 & 54\,\% & 91\,\% \\
    DE-23 & 90\,\% & 75\,\% & 90\,\% && AE-23 & 61\,\% & 89\,\% \\
    DE-24 & 95\,\% & 94\,\% & 92\,\% && AE-24 & 77\,\% & 90\,\% \\
    \midrule
    DE-2ALL & 91\,\% & 85\,\% & 88\,\% && AE-2ALL & 62\,\% & 87\,\% \\
    \bottomrule
    \end{tabular}

\end{table}

\subsection{Classification Accuracy Convergence}
\label{sec:convergence-analysis}

Whereas the results reported for the classification so far always refer to the maximum number of notification timing sequences available for all classes, we are also interested in how many samples are actually required for an accurate classification.
To this end, we repeatedly run four specific classifications representing different classification tasks with increasing numbers of notification timing samples.
We start with \num{10} samples per class and increase this number in steps of \num{10} until we reach \num{300} or the maximum number of available samples for all classes (if it is lower than \num{300}).
Figure~\ref{fig:model_convergence} illustrates the results of these evaluations.
We include (a)~the receiver country classification based on the first round of measurements, two classifications of three WiFi locations, both (b)~device-specific in Germany (device \emph{DE-23}) and (c)~cross-device in the UAE (referred to as \emph{AE-2ALL}), and (d)~a receiver network classification for one of the devices (\emph{DE-22}) in Germany.
Whereas the overall classification accuracy is varying for smaller sample sizes, there are only minor improvements for more than around \num{100} sequences of \num{5} delivery confirmation timings.
This observation seems to hold for all three messengers and across the different classification tasks.% under consideration.
Thus, we can already reach considerable classification results with sample sizes of around \num{100} delivery confirmation timing sequences per class~--~for some cases, \eg, the network connection detection, even with lower sample sizes.

\begin{figure*}[tb]
\centering
\includegraphics[width=\textwidth]{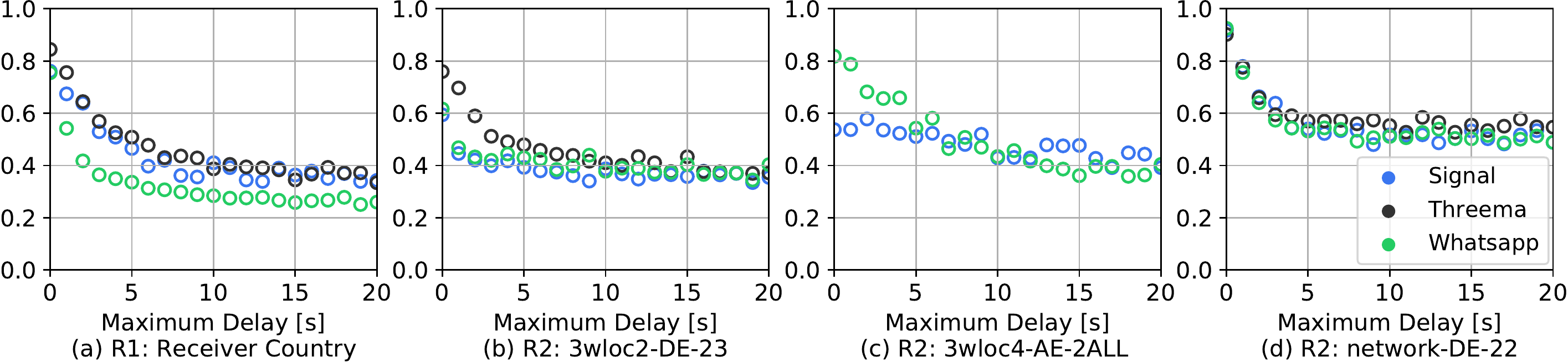}
\caption{Overall accuracy of four different classification tasks with increasing random delays (x-axis) added to message delivery confirmation timings. For higher delays, the accuracy approximates the chance of randomly guessing the receiver's location.~\label{fig:cm_eval}} 
\end{figure*}

\subsection{Experimental Factors}
\label{sec:factors}
While we are mostly interested in differences between receiver characteristics such as their location or network connection type, there are many dynamic features that can influence the RTTs of delivery confirmations, including network, device, and server characteristics.
We now carefully discuss how such features are reflected in our measurements, and to what extent they can affect our experiments.

\paragraph*{Network Characteristics}
Varying network loads, both in terms of general Internet traffic and messenger use, may affect the time required to send a message and receive the confirmations.
However, such circumstances cannot be influenced by our setup.
In general, network loads are mostly varying depending on the time of day, with higher loads during mornings and evenings~\cite{feldmann20:lockdown-effect-implications,trevisan20:five-years-edge}.
Since we continuously collect data for at least one week per receiving device and location, all relevant load levels should be covered by our measurements.
When looking into our timing dataset, we do not observe large deviations or suspicious patterns depending on the time of day.
Thus, the influence of network load on our dataset should be negligible.
Figure~\ref{fig:r1_timhod_de11} in Appendix~\ref{app:timings} serves as an example and shows detailed distributions of $RTT_{M,R}$ for messages sent from device~\emph{DE-11} to receiving devices in 
their respective countries per hour of day.

Timings may also depend on the routes taken between sender, messenger server, and receiver, which could vary depending on the provider of the devices' Internet connections, making WiFi locations easier distinguishable when different connection providers are involved.
In our measurements in Germany, only locations DE-C and DE-E were using the same connection provider but, unfortunately, our dataset does not include measurements of the same device in both locations.
In the UAE, all Internet connections were provided by the same operator, with timings being fairly distinguishable (\eg, \SI{82}{\percent} accuracy for \emph{2wloc1-AE-22}).
However, our dataset is too small, to adequately measure the effect of using the same provider at multiple locations vs. using different ones.

\paragraph*{Device Behavior}
During our measurements, receiving devices were idling at each location while receiving messages.
This comprises a limitation of our setup, since active interaction with the devices and  parallel processes may affect the timings we measure while sending messages, with potential consequences for classification accuracy.

To overcome this issue, we conducted additional experiments over one week sending messages to one author's private smartphone while it was in everyday use and continuously recorded its network connection type (\ie, WiFi or mobile data). 
We then used the data to predict its network connection following the procedures described in Section~\ref{sec:classification}.
Classification reaches overall accuracy of \SI{82}{\percent} for Signal, \SI{80}{\percent} for Threema, and \SI{74}{\percent} for WhatsApp.
These numbers are fairly lower than the ones in our original and fully controlled setup (\cf~Table~\ref{tab:msg_timings-accuracy-network}) and shows that the threat vector still persists in a realistic usage profile, although with lower accuracy.

\paragraph*{Server Behavior}
Through the experiments, the sender devices were connected to different servers when sending WhatsApp messages.
We only consider WhatsApp here, since Threema only has one server location and Signal's actual infrastructure remains unclear.
While the same sender connected to up to \num{34} different WhatsApp IP addresses (\emph{AE-21}), \num{3} servers (\num{4} for \emph{DE-21}, respectively) make up at least \SI{95}{\percent} of connections used when sending messages.
Additionally, server selection follows similar distributions for all receiver locations.
Thus, the selected server should have little unintended influence on our measurements.
While our data does not contain meaningful differences in round-trip times depending on the selected server, it may be possible that strategic server selection could help the attacker (\eg, by locally changing DNS resolution) to make timings better distinguishable, \ie, further improve classification accuracy.
We leave the required data collection and evaluation an open task for future work.

\section{Countermeasures}
\label{sec:countermeasures}
We now shed light on possible countermeasures that can be applied to make the receiver location classification harder to better protect clients' location privacy.
We consider countermeasures on the messenger's and on the user's side.

\subsection{Randomizing Delivery Confirmation Times}
Since timing measurements are a noisy source of information used for the attack, randomly delaying the delivery confirmation might be a viable solution to make timings to receivers in different locations harder to distinguish.
While adding random delays must be implemented by messenger providers to come into effect, we can evaluate the impact of such a mechanism through a simulation based on the timing data we collected.

Timings can be perturbed by adding a delay sampled uniformly at random \emph{between 0 and a specific maximum delay}.
We systematically repeat the evaluation of the same four classification tasks (\cf~Section~\ref{sec:convergence-analysis}) and increase the maximum delay in every iteration by \SI{1}{second} from \SI{0}{s} to \SI{20}{s}.
Our goal is to find a threshold value that is sufficient to make the delivery confirmation timings to receivers in different locations indistinguishable.
In addition, the maximum delay should be as small as possible to keep the impact on user experience low.

Figure \ref{fig:cm_eval} shows the overall accuracy values for four classification tasks with maximum random delays between \SI{0}{s} and \SI{20}{s}.
We selected the same classification tasks as for the classification accuracy convergence analysis, again, to cover different types of classifications (\cf~Section~\ref{sec:convergence-analysis}).
A maximum delay of \SI{0}{s} corresponds to the original classification results. 
When we increase the maximum delay, the overall classification accuracy continuously decreases and approximates the chance of randomly guessing the location, which depends on the number of location candidates.
Depending on the classification task, the random guessing accuracy is reached for a maximum delay of between \SI{5}{s} and \SI{10}{s}, as for example for determining the network connection of receiving device \emph{DE-22} (\cf Figure~\ref{fig:cm_eval}d).
Messenger servers randomly delaying delivery confirmations by up to \SI{6}{s} seems to be sufficient to render the timings indistinguishable and, thus, to disable the timing side channel in message delivery confirmations. 
We emphasize that there is a graceful degradation of accuracy with increasing delays~--~introducing maximum delays of as little as 1 or 2 seconds will already have a positive and measurable impact on users' location privacy under our attack.

If and to what extent the maximum delay can be further decreased or even flexiblized, \eg, different delays for different groups of contacts, or depending on dynamic parameters should be subject to extensive further evaluations.
The best option from a user perspective would actually be the possibility to disable sending (and receiving) delivery confirmations at all~--~exactly as it is already offered for \emph{read receipts} (verbatim a privacy option) in all messengers we analyzed in this paper.

\subsection{User-side countermeasures}
Users' means to reduce the effects of the timing side channel are limited, since delivery confirmations cannot be turned off in the messengers we analyzed~--~randomly delaying these timings can only be applied by the messenger providers.
However, the use of VPN services or Tor routing all traffic through dedicated servers at distant and changing geographical locations may be a promising mitigation strategy that can be applied by users.
The overhead of additional servers may perturb the delivery notifications in a similar fashion like adding random delays.

We run a small additional experiment to get a preliminary estimate of the effects of using a VPN as a countermeasure. 
To this end, we send messages to one receiver phone (\emph{DE-23}) in one location (\emph{DE-B}) both on WiFi and cellular Internet connections~--~in both cases connected to a US-based VPN server provided by a commercial VPN provider.
Whereas without VPN, the network connection of this device can be distinguished with up to \SI{90}{\percent} accuracy (\cf~Table~\ref{tab:msg_timings-accuracy-network}, classifications perform worse when using a VPN.
For Threema (\SI{51}{\percent}) and WhatsApp (\SI{62}{\percent}), performance is hardly better than random guessing (\SI{50}{\percent}).
However, for Signal, we reach a surprisingly high overall accuracy of \SI{77}{\percent}.
When repeating the same small experiment with using Tor instead of a VPN, WiFi and cellular connections can be distinguished better (Signal: \SI{72}{\percent}, Threema: \SI{58}{\percent}, WhatsApp: \SI{82}{\percent}).

Without investigating these issues more systematically, we can only speculate about the reasons.
One explanation could be that Signal's servers are US-based and, therefore, the routing overhead introduced by using the VPN server is too small to adequately perturb timings.
For the case of Tor, the set of circuits selected in either sample may have biased the comparably small sets of timings we measured. 
However, since conclusive statements require more systematic and extensive measurements to allow a thorough evaluation, we leave this issue an open task for future work.

Since users' means to perturb timings and, thus, to disable the side channel seem ineffective in practice, another option could be to totally block delivery confirmations, \eg, by filtering the related packets based on their size out of their local network traffic by means of a firewall.
While this might be a viable solution for technically adept users or in specifically security-sensitive use cases, it does, however, not apply to the vast majority of the 2~billion WhatsApp users.
\section{Related Work}
\label{sec:rw}

\subsubsection*{Security of Messengers}
A systematization of knowledge by Unger~\etal~\cite{unger15:sok:-secure-messaging} provides an extensive overview of security features in many instant messaging applications.
Similarly, also other studies have analyzed security features of specific subsets of messengers and their cryptographic foundations~\cite{aggarwal18:security-aspect-instant,herzberg16:johnny-finally-encrypt,johansen17:comparing-implementations-secure}, protocols~\cite{cohn17:formal-analysis-signal,frosch16:secure-textsecure,kobeissi17:automated-verification,rosler18:more-less-end}, or exploited specific features such as contact discovery to crawl millions of American phone numbers~\cite{hagen21:numbers-large-scale}.

Additionally, the analysis of encrypted messenger traffic has served as a side channel to identify the language used in iMessage conversations~\cite{coull14:traffic-analysis-encrypted}, specific user actions in KakaoTalk~\cite{park15:encryption-enough-inferring}, and users in various messengers~\cite{bahramali20:practical-traffic-analysis}.
Our paper complements such works in providing empirical evidence for a similar side channel under real-world conditions.
Different from these works, though, our attack is conducted from one participant and directed at a specific target.

Despite such attacks, messengers specifically designed to improve the privacy of contact discovery~\cite{kales19:mobile-contact-discovery} or to resist traffic analysis~\cite{hooff15:vuvuzela-resistant-traffic-analysis} are, however, not widely in use.

\subsubsection*{Analysis of Timings and Internet Traffic}
There is a large body of work studying the connection between timings and distances and taking into account the Internet topology for the purpose of  localization~\cite{candela19:using-ripe-atlas,du20:ripe-ipmap-active,katz-bassett06:towards-geolocation-using,kohls22:verloc-verifiable-localization} and distance bounding~\cite{avoine18:security-distance-bounding,mauw18:distance-bounding-protocols,peeters18:sonar-detecting-redirection,rasmussen10:realization-distance-bounding} on the Internet.
Our work is different, in that we do not directly relate timings to traveled distances, but instead use recurring timing characteristics to re-identify previously seen, expected locations.
Similarly, traffic analysis~\cite{danezis04:traffic-analysis-continuous,shmatikov06:timing-analysis-low} is regularly used to analyze encrypted network traffic in various other domains. 
Purposes include website fingerprinting~\cite{kohls19:lost-traffic-encryption,panchenko18:website-fingerprinting-internet,rimmer18:automated-website-fingerprinting} or deanonymizing users and their end-to-end connections in anonymity networks such as Tor~\cite{biryukov13:trawling-hidden-services,geddes13:balancing-performance-with,kwon15:circuit-fingerprinting-attacks,nasr18:deepcorr-strong-flow,nasr17:compressive-traffic-analysis,rimmer22:trace-oddity-methodologies,schnitzler21:we-built-this-circuit} with the most recent ones technically reaching accuracy of up to \SI{100}{\percent} using deep learning techniques.

\section{Conclusion}

We presented a novel timing side-channel in popular instant messengers, allowing to distinguish different receivers and their locations by sending them instant messages.
We have demonstrated how measuring the time between sending a message and receiving the notification that the message has been delivered enables clients to spy on each other, \eg, to determine whether or not they are at their usual location.
While making use of this side channel is mostly limited to people who are in each others' contact lists and have already started a conversation before, it yet comprises an unexpected and privacy-infringing act with low technical requirements that is equally hard to detect and to mitigate for a potential victim.

\section*{Acknowledgement}
This research was supported in parts by the Research Center Trustworthy Data Science and Security, one of the Research Alliance centers within UA Ruhr, and by the Center for Cyber Security at NYU Abu Dhabi. The authors would like to thank Marvin Kowalewski, Leona Lassak, Philipp Markert, Sarah Pardo, and Lena Schnitzler for their help with data collection.

%\balance
\bibliographystyle{plain}
\bibliography{msgloc}

%\cleardoublepage

%\onecolumn
%\newpage
\appendix

\subsection{Parameter Tuning Configuration Details}
\label{sec:appendix-nn-tuning}
The tuned parameters for each neural network type are listed below, the best performing configuration is highlighted in \textbf{bold}.

\paragraph*{Convolutional Neural Network (CNN)}
\begin{itemize}
    \item Activation function: tanh, \textbf{relu}
    \item Optimizer: SGD, \textbf{Adam}, RMSProp
    \item Dropout rate: 0, \textbf{0.1}, 0.2, 0.3
    \item Number of epochs: 20, 30, 40, 50, \textbf{60}
    \item CNN input filters: 8, 16, \textbf{32}, 64
    \item Number of fully connected layers: 1, \textbf{2}, 3, 4, 5
    \item Number of neurons on fully connected layers: \textbf{50}, 100, 200, 500
\end{itemize}

\paragraph*{Long Short-Term Memory Recurrent Neural Network (LSTM-RNN)}
\begin{itemize}
    \item Activation function: tanh, Sigmoid, \textbf{relu}
    \item Optimizer: SGD, \textbf{Adam}, RMSProp
    \item Dropout rate: \textbf{0}, 0.1, 0.2, 0.3
    \item Number of epochs: 20, 30, 40, 50, \textbf{60}
    \item Number of LSTM layers: 1, 2, \textbf{3}, 4, 5
    \item Number of LSTM units: 50, \textbf{100}, 200, 500
\end{itemize}

\paragraph*{Stacked Denoising Autoencoder (SDAE)}
\begin{itemize}
    \item Activation function: \textbf{tanh}, Sigmoid, relu
    \item Optimizer: SGD, Adam, \textbf{RMSProp}
    \item Dropout rate: \textbf{0}, 0.1, 0.2, 0.3
    \item Number of epochs: 20, 30, 40, \textbf{50}, 60
    \item Number of encoding layers: \textbf{1}, 2, 3
\end{itemize}

\onecolumn

\subsection{Messenger Infrastructures}
\label{sec:appendix-worldmap}

\begin{figure}[h!]%{\textwidth}
\centering

\includegraphics[width=0.95\textwidth]{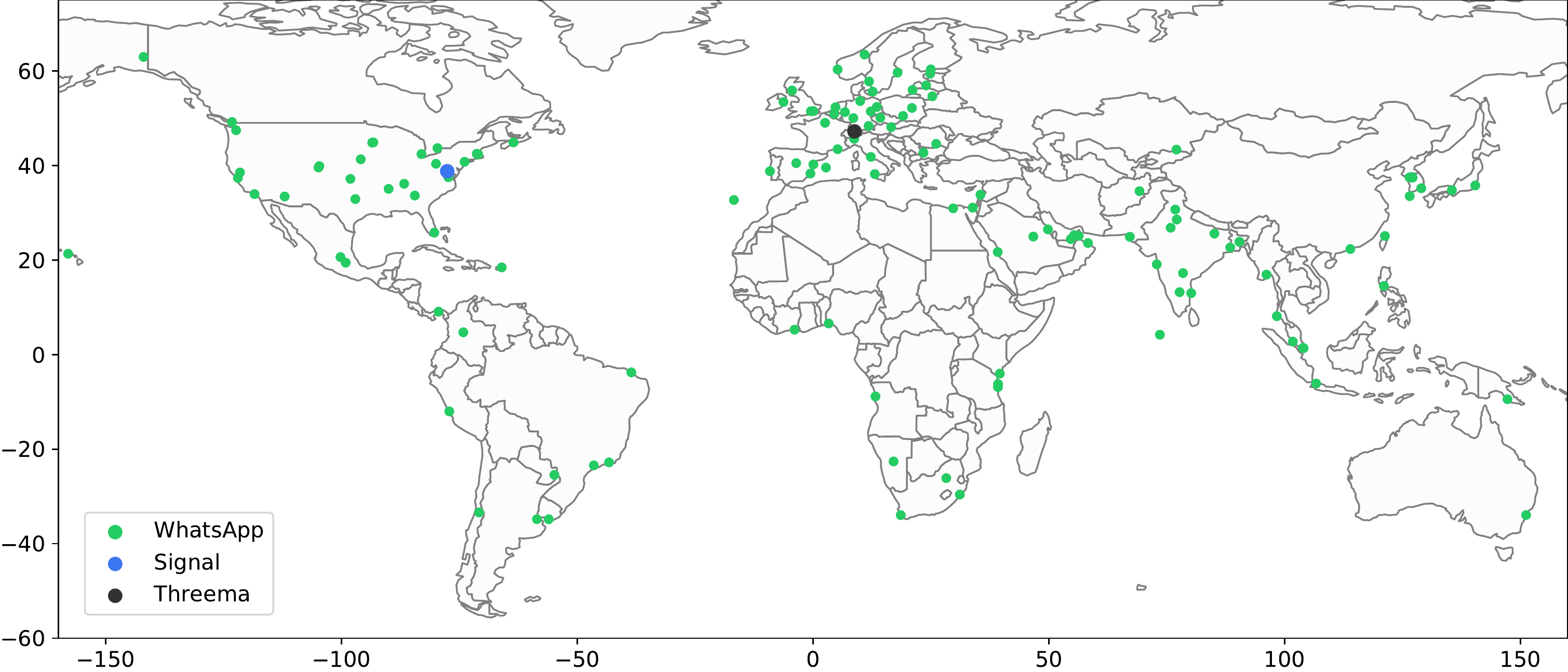}

\captionof{figure}{Locations of Signal, Threema, and WhatsApp servers around the world. Signal is located at the US east coast, Threema is hosted in Switzerland, and WhatsApp instances are widely distributed across all continents.~\label{fig:worldmap}} 
\end{figure}

\subsection{Detailed Timing Data}
\label{app:timings}

\begin{minipage}{\textwidth}

\emph{1) Round 1 (Hour of day)}

\begin{center}
\includegraphics[width=\textwidth]{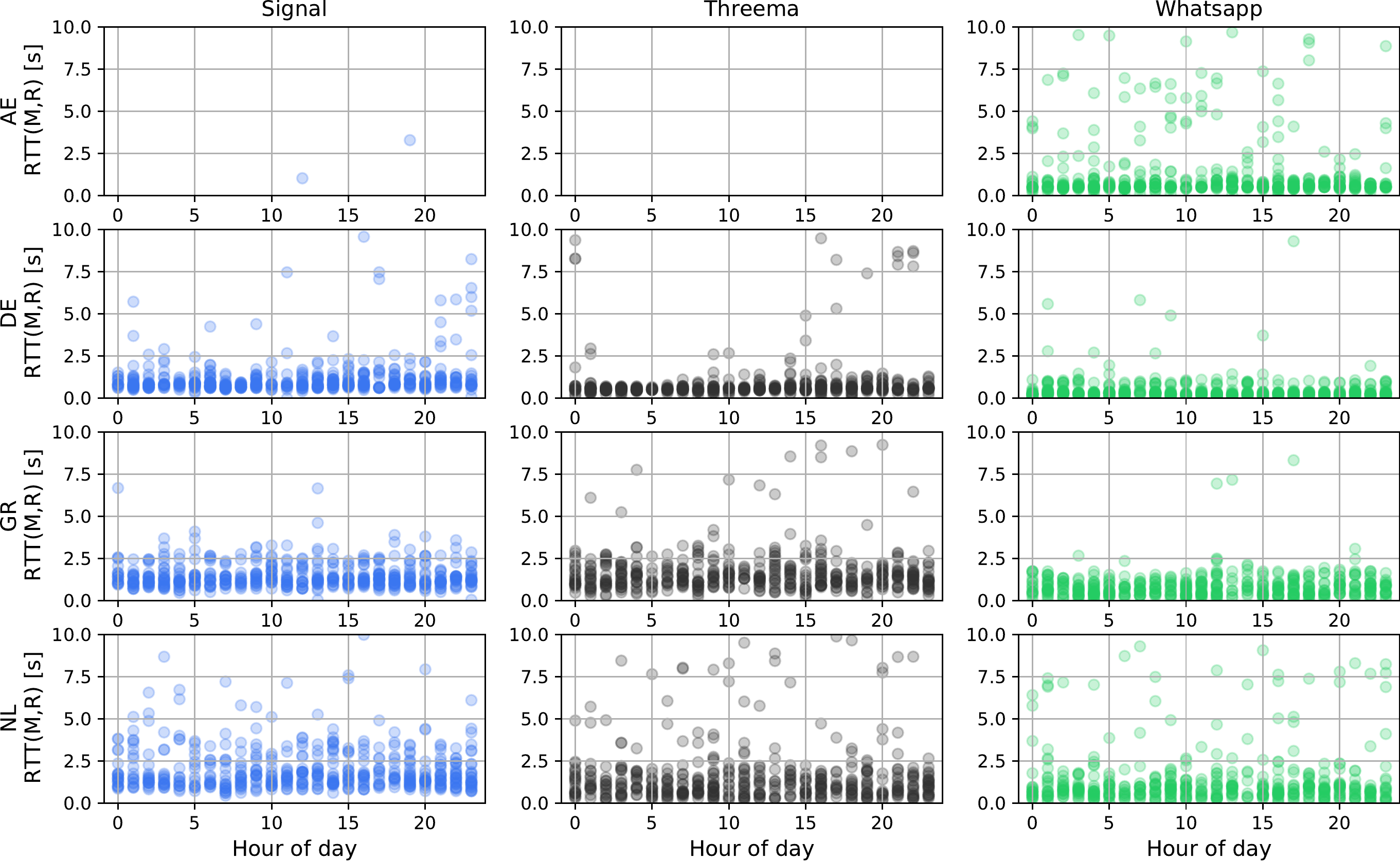}
\captionof{figure}{Distributions of $RTT_{M,R}$ of messages sent from device \emph{DE-11} to receivers in different countries per hour of day.~\label{fig:r1_timhod_de11}} 
\end{center}

\end{minipage}

\begin{minipage}{\textwidth}
\centering

\includegraphics[width=\textwidth]{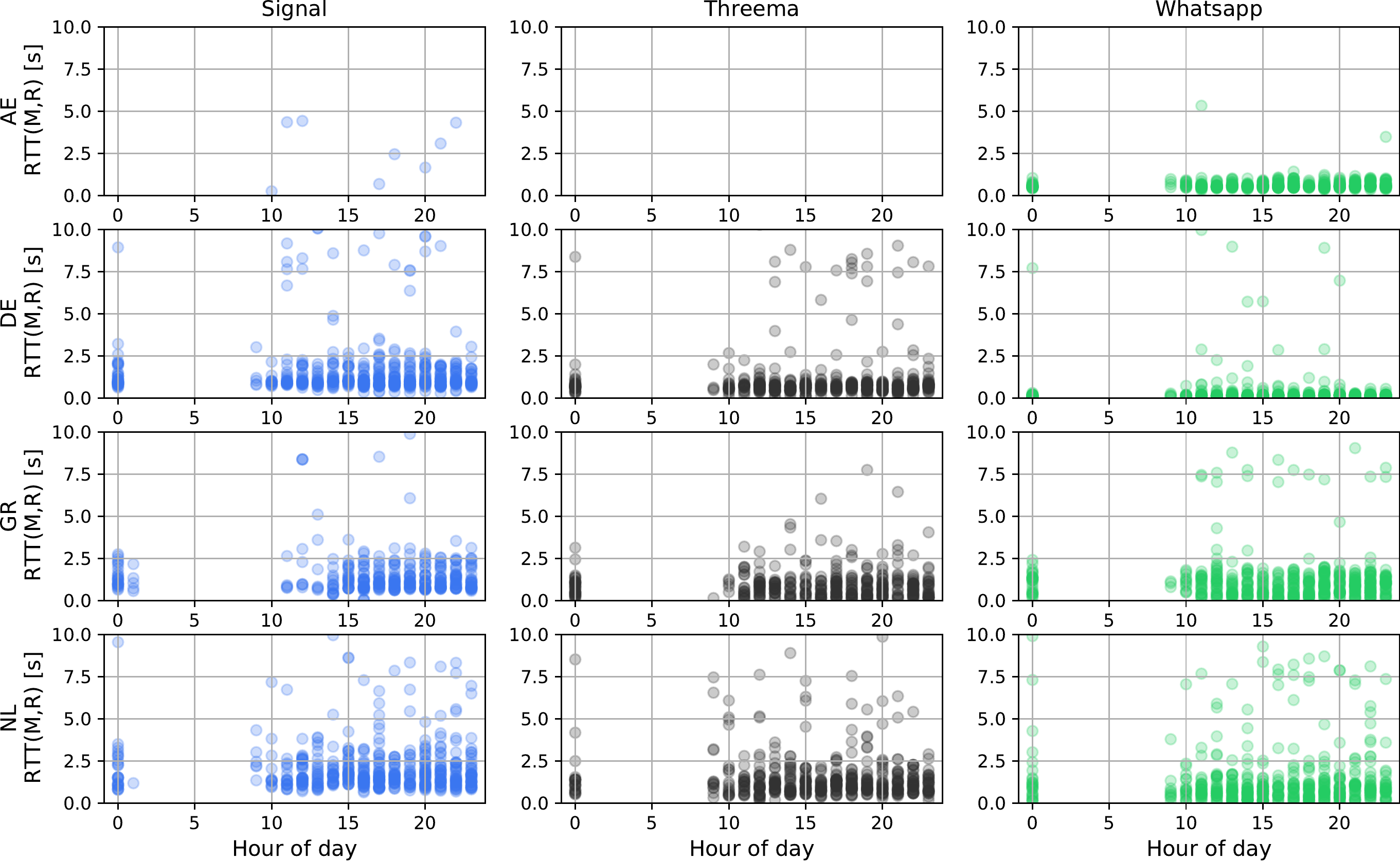}
\captionof{figure}{Distributions of $RTT_{M,R}$ of messages sent from device \emph{DE-12} to receivers in different countries per hour of day.~\label{fig:r1_timhod_de12}} 

\includegraphics[width=\textwidth]{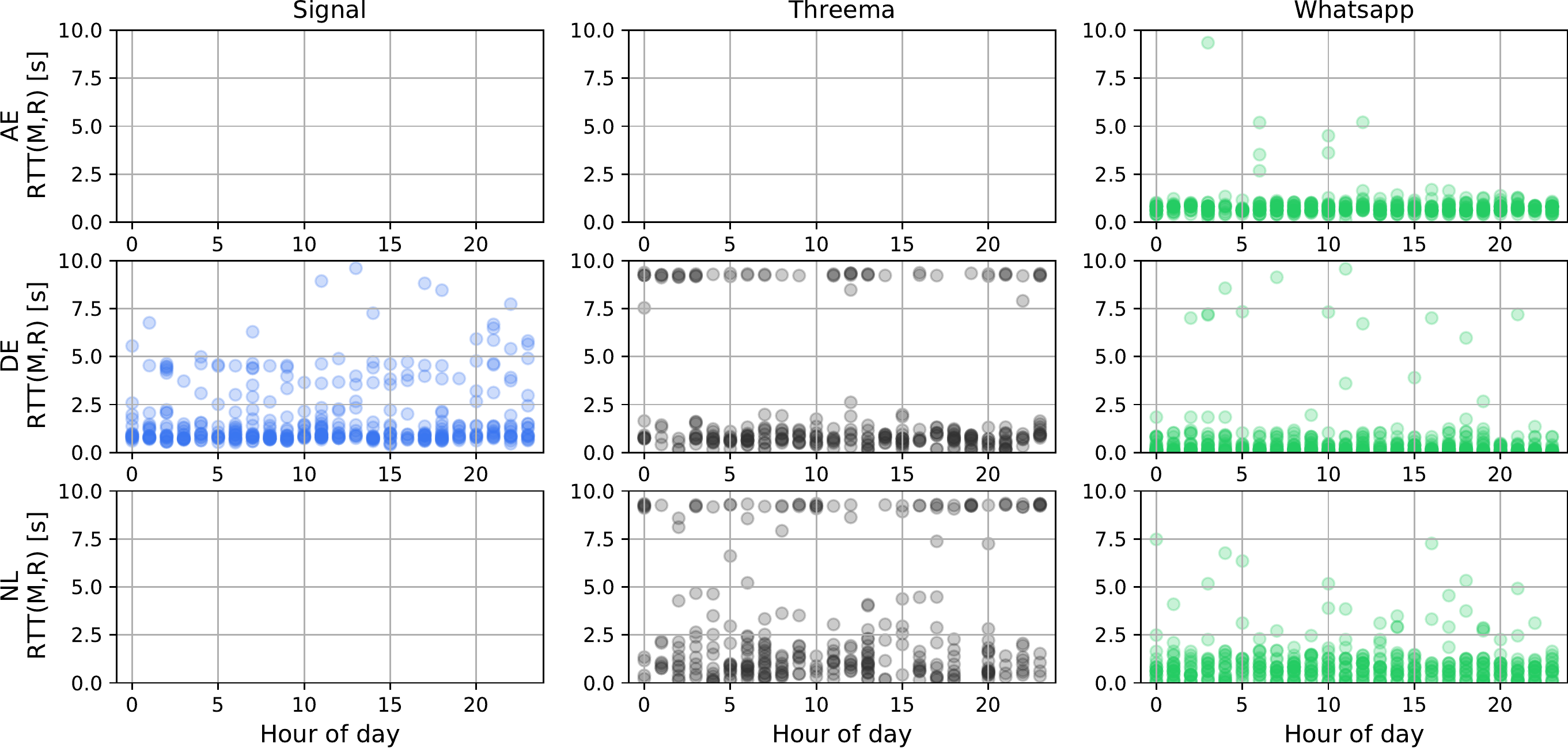}
\captionof{figure}{Distributions of $RTT_{M,R}$ of messages sent from device \emph{GR-11} to receivers in different countries per hour of day.~\label{fig:r1_timhod_gr11}} 

\end{minipage}

\begin{minipage}{\textwidth}

\begin{center}
\includegraphics[width=\textwidth]{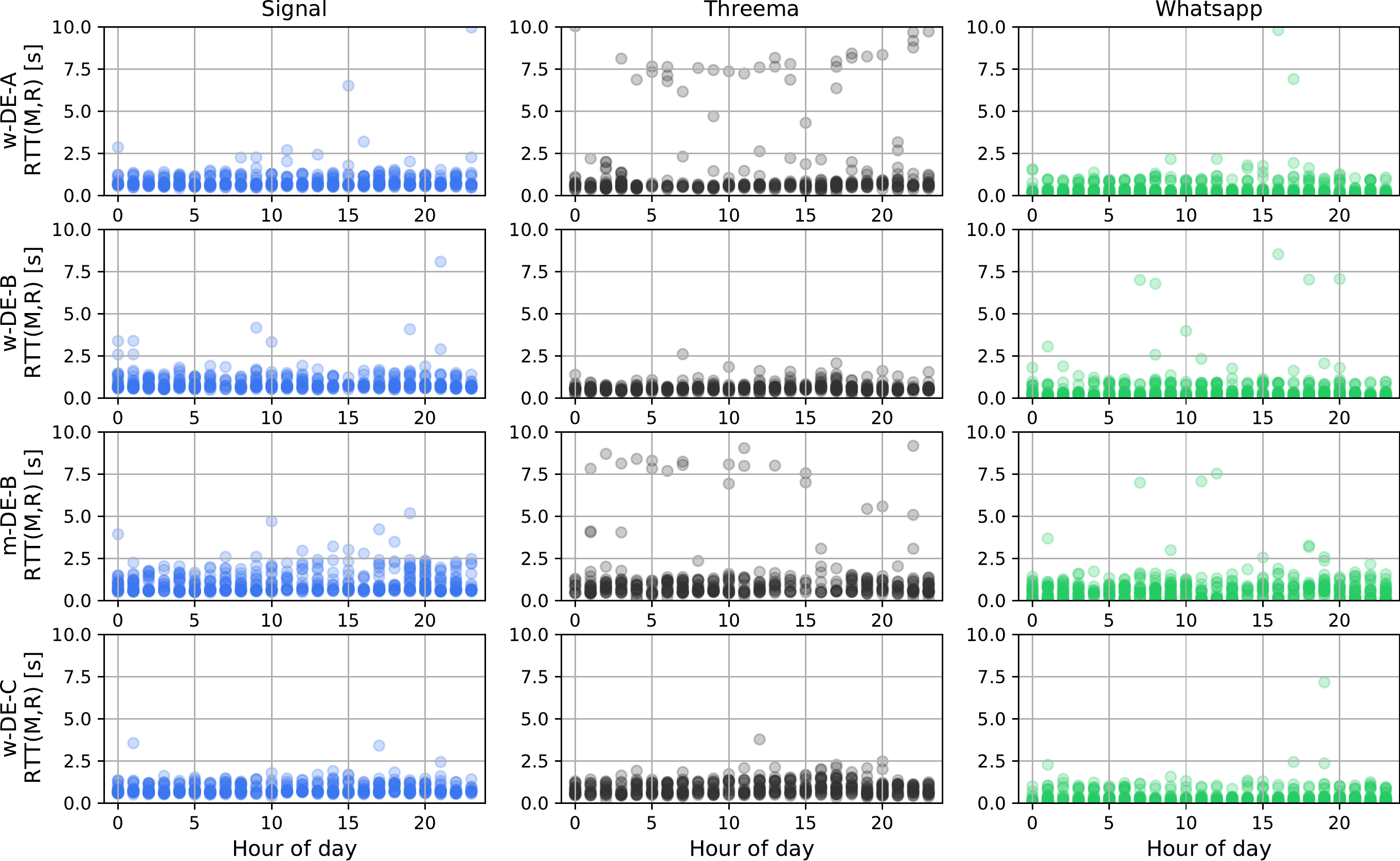}
\captionof{figure}{Distributions of $RTT_{M,R}$ of messages sent to device \emph{DE-22} at each location (rows) per hour of day.~\label{fig:r2_timhod_de22}} 
\end{center}

\vspace{0.1cm}
\emph{2) (Hour of day)}

\begin{center}
\includegraphics[width=\textwidth]{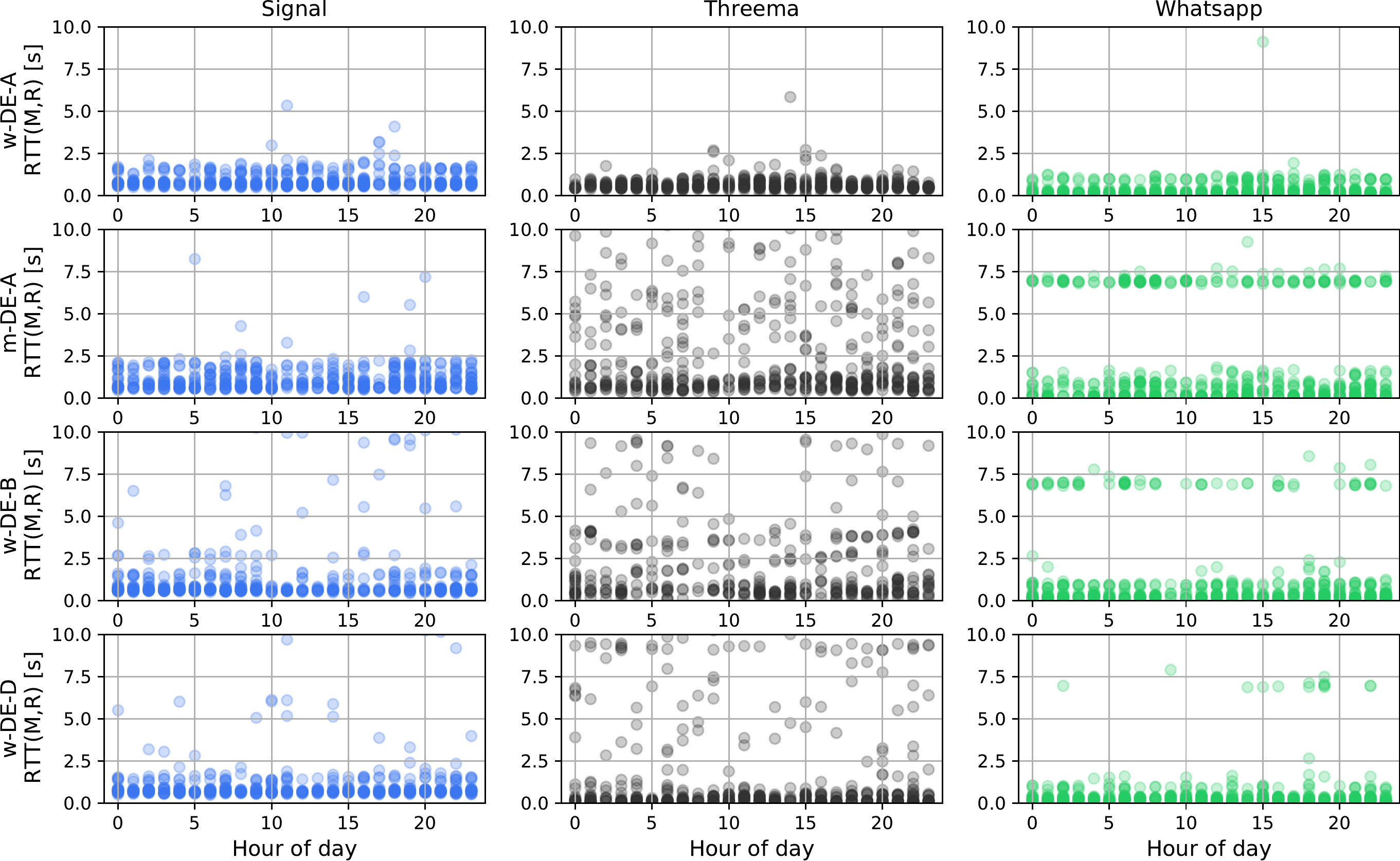}
\captionof{figure}{Distributions of $RTT_{M,R}$ of messages sent to device \emph{DE-23} at each location (rows) per hour of day.~\label{fig:r2_timhod_de23}} 
\end{center}

\end{minipage}

\begin{minipage}{\textwidth}
\centering

\includegraphics[width=\textwidth]{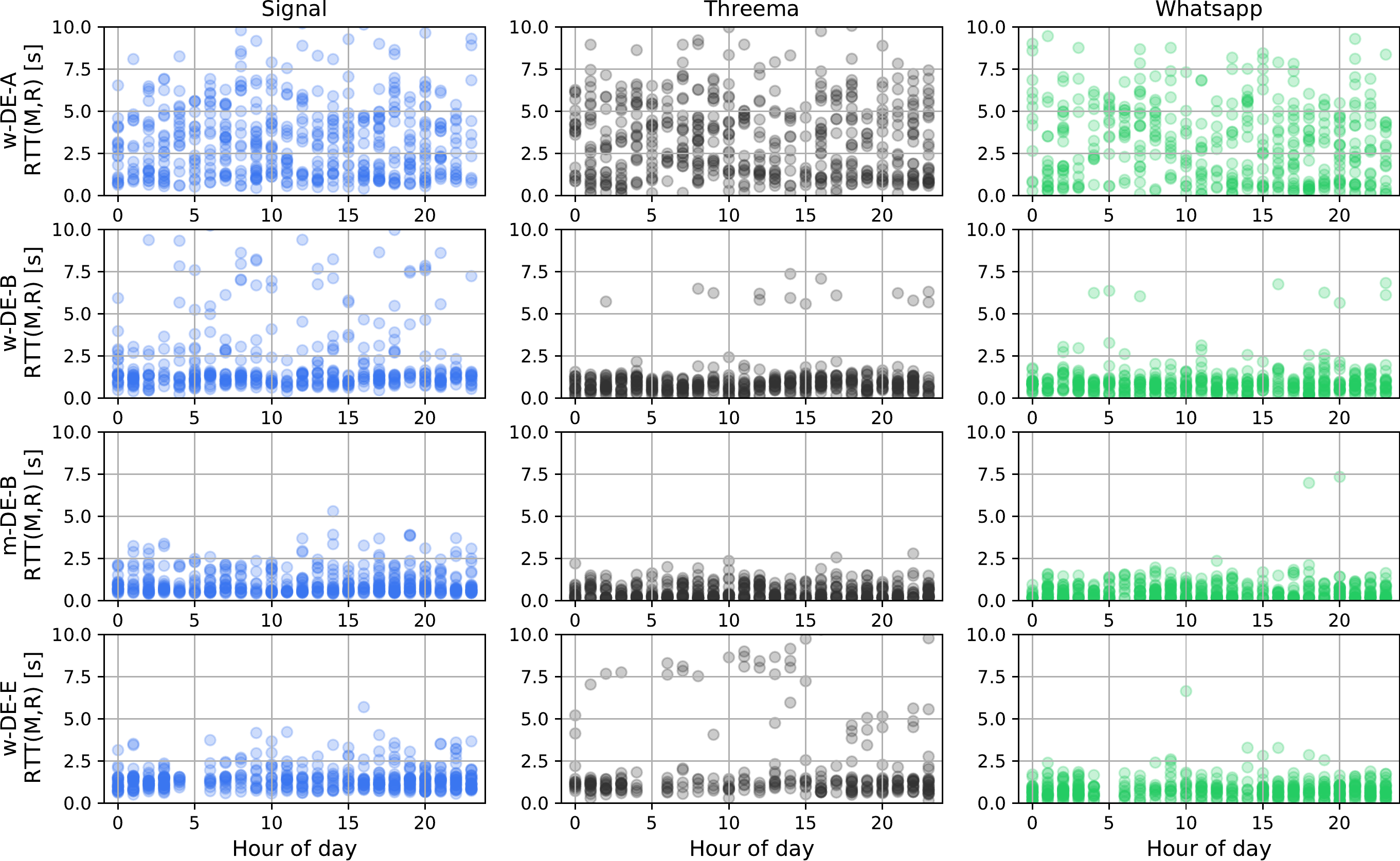}
\captionof{figure}{Distributions of $RTT_{M,R}$ of messages sent to device \emph{DE-24} at each location (rows) per hour of day.~\label{fig:r2_timhod_de24}} 

\includegraphics[width=\textwidth]{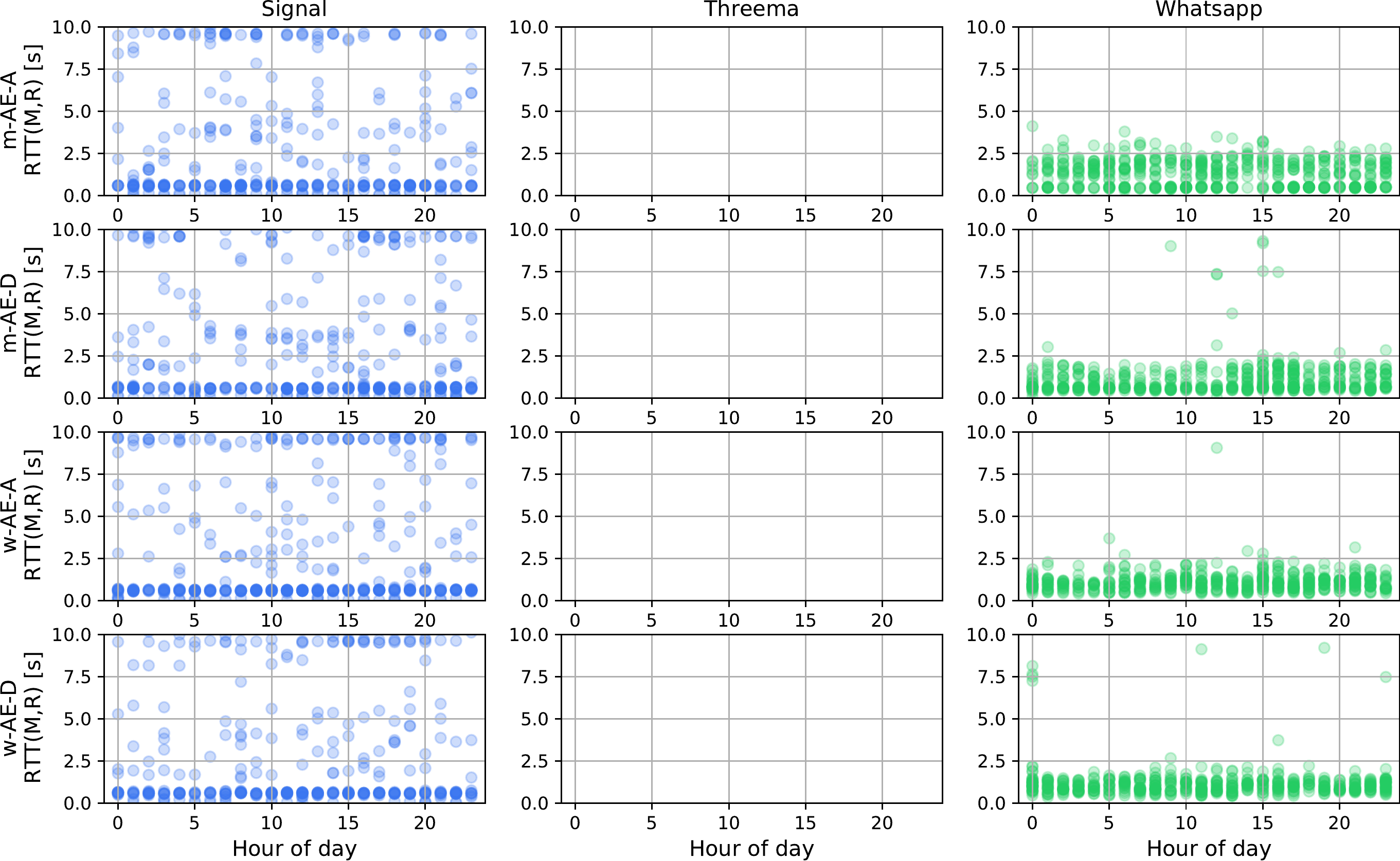}
\captionof{figure}{Distributions of $RTT_{M,R}$ of messages sent to device \emph{AE-22} at each location (rows) per hour of day.~\label{fig:r2_timhod_ae22}} 

\end{minipage}

\begin{minipage}{\textwidth}
\centering

\includegraphics[width=\textwidth]{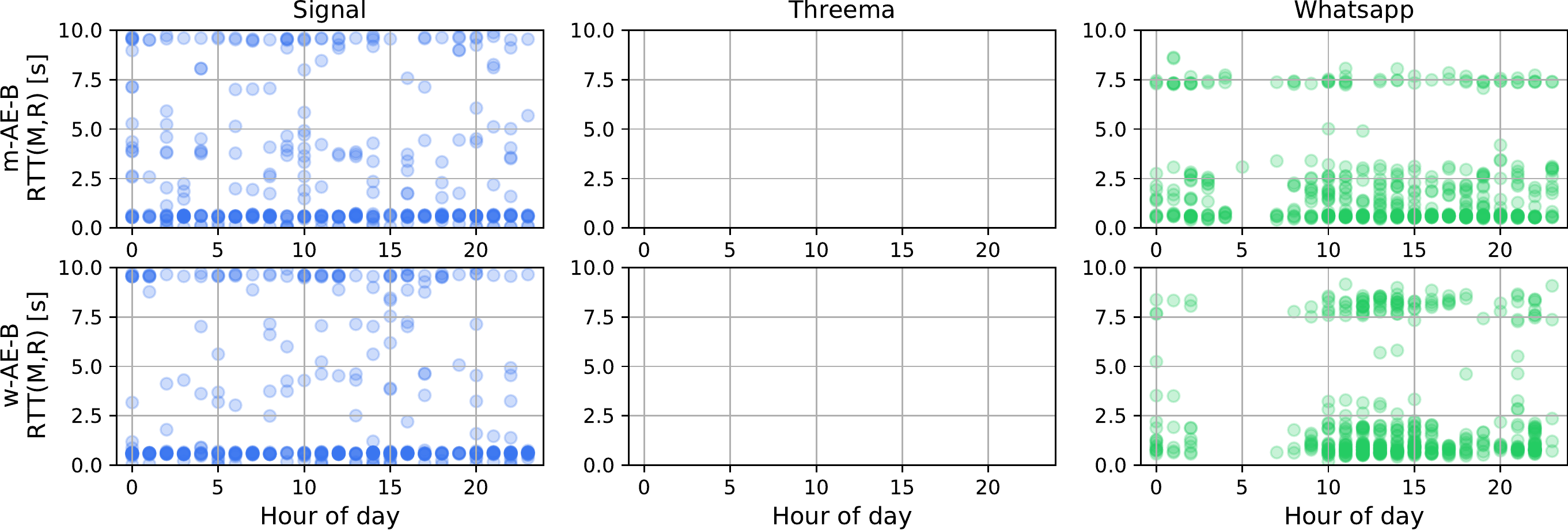}
\captionof{figure}{Distributions of $RTT_{M,R}$ of messages sent to device \emph{AE-23} at each location (rows) per hour of day.~\label{fig:r2_timhod_ae23}} 

\includegraphics[width=\textwidth]{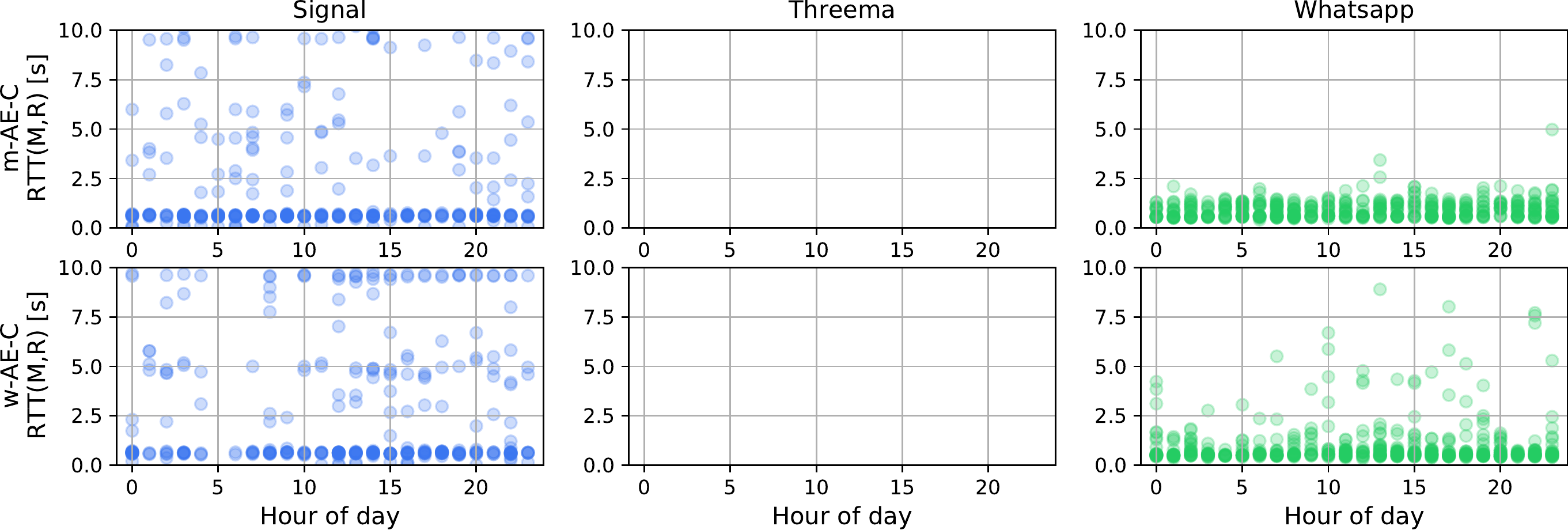}
\captionof{figure}{Distributions of $RTT_{M,R}$ of messages sent to device \emph{AE-24} at each location (rows) per hour of day.~\label{fig:r2_timhod_ae24}}

\end{minipage}

\subsection{Detailed Classification Results}
\label{sec:appendix-detailed}

\subsubsection{Round 1}

\footnotesize

\begin{center}
\begin{longtable}{llrrrrr|rrrrr|rrrrr}
\caption{\label{tab:details-r1}Detailed classification results for the first round of measurements. Precision values for each class and overall classification accuracy values. Five values per messenger represent different notification sequence lengths.}\\

\toprule \textbf{Classification Task} & \textbf{Receiver Loc.} & \multicolumn{5}{c}{\textbf{Signal}} & \multicolumn{5}{|c|}{\textbf{Threema}} &  \multicolumn{5}{c}{\textbf{Whatsapp}}\\
&& 1&2&3&4&5&1&2&3&4&5&1&2&3&4&5\\
\midrule
\endfirsthead

\multicolumn{17}{c}%
{{\bfseries \tablename\ \thetable{} -- continued from previous page}} \\
\midrule  \textbf{Classification Task} & \textbf{Receiver Loc.} & \multicolumn{5}{c}{\textbf{Signal}} & \multicolumn{5}{|c|}{\textbf{Threema}} &  \multicolumn{5}{c}{\textbf{Whatsapp}}\\ 
&& 1&2&3&4&5&1&2&3&4&5&1&2&3&4&5\\
\midrule
\endhead

\midrule \multicolumn{17}{r}{{Continued on next page}} \\ \midrule
\endfoot

\hline \hline
\endlastfoot

\multicolumn{17}{l}{\emph{Two countries measured with sender DE-11}}\\
\midrule
DE11-2countries1&Overall Acc.&0.81&0.83&0.85&0.85&0.91&0.73&0.81&0.84&0.86&0.92&0.77&0.85&0.89&0.9&0.91\\
DE11-2countries1&DE&0.77&0.81&0.88&0.86&0.89&0.83&0.82&0.83&0.89&0.91&0.84&0.83&0.88&0.89&0.9\\
DE11-2countries1&NL&0.84&0.86&0.82&0.85&0.92&0.63&0.79&0.84&0.84&0.93&0.7&0.87&0.91&0.91&0.93\\
\midrule
DE11-2countries2&Overall Acc.&&&&&&&&&&&0.82&0.85&0.85&0.86&0.95\\
DE11-2countries2&AE&&&&&&&&&&&0.88&0.9&0.87&0.91&0.96\\
DE11-2countries2&DE&&&&&&&&&&&0.76&0.8&0.84&0.8&0.94\\
\midrule
DE11-2countries3&Overall Acc.&0.76&0.85&0.87&0.88&0.89&0.87&0.9&0.92&0.95&0.96&0.77&0.83&0.84&0.86&0.9\\
\end{longtable}
\end{center}
\subsubsection{Round 2 (United Arab Emirates)}

\footnotesize

\begin{center}
\begin{longtable}{llrrrrr|rrrrr}
\caption{\label{tab:details-r2ae}Detailed classification results for the second round of measurements in the UAE. Precision values for each class and overall classification accuracy values. Five values per messenger represent different notification sequence lengths.}\\

\toprule \textbf{Classification Task} & \textbf{Receiver Loc.} & \multicolumn{5}{c}{\textbf{Signal}} & \multicolumn{5}{c}{\textbf{Whatsapp}}\\
&& 1&2&3&4&5&1&2&3&4&5\\
\midrule
\endfirsthead

\multicolumn{12}{c}%
{{\bfseries \tablename\ \thetable{} -- continued from previous page}} \\
\midrule  \textbf{Classification Task} & \textbf{Receiver Loc.} & \multicolumn{5}{c}{\textbf{Signal}} &  \multicolumn{5}{c}{\textbf{Whatsapp}}\\ 
&& 1&2&3&4&5&1&2&3&4&5\\
\midrule
\endhead

\midrule \multicolumn{12}{r}{{Continued on next page}} \\ \midrule
\endfoot

\hline \hline
\endlastfoot

\multicolumn{12}{l}{\emph{Two cellular locations}}\\
\midrule
2mloc1-AE-2ALL&Overall Acc.&0.5&0.5&0.52&0.48&0.51&0.65&0.85&0.85&0.85&0.91\\
2mloc1-AE-2ALL&m-AE-A&0.6&0.54&0.47&0.42&0.49&0.85&0.88&0.87&0.87&0.92\\
2mloc1-AE-2ALL&m-AE-D&0.4&0.45&0.57&0.54&0.53&0.46&0.81&0.82&0.84&0.89\\
\midrule
2mloc1-AE-22&Overall Acc.&0.49&0.49&0.5&0.52&0.5&0.63&0.82&0.87&0.85&0.9\\
2mloc1-AE-22&m-AE-A&0.68&0.51&0.44&0.51&0.44&0.79&0.87&0.91&0.88&0.93\\
2mloc1-AE-22&m-AE-D&0.3&0.47&0.57&0.52&0.56&0.47&0.77&0.82&0.83&0.87\\
\midrule
2mloc2-AE-2ALL&Overall Acc.&0.49&0.51&0.49&0.52&0.49&0.73&0.89&0.89&0.94&0.92\\
\end{longtable}
\end{center}
\subsubsection{Round 2 (Germany)}

\footnotesize

\begin{center}
\begin{longtable}{llrrrrr|rrrrr|rrrrr}
\caption{\label{tab:details-r2de}Detailed classification results for the second round of measurements in Germany. Precision values for each class and overall classification accuracy values. Five values per messenger represent different notification sequence lengths.}\\

\toprule \textbf{Classification Task} & \textbf{Receiver Loc.} & \multicolumn{5}{c}{\textbf{Signal}} & \multicolumn{5}{|c|}{\textbf{Threema}} &  \multicolumn{5}{c}{\textbf{Whatsapp}}\\
&& 1&2&3&4&5&1&2&3&4&5&1&2&3&4&5\\
\midrule
\endfirsthead

\multicolumn{17}{c}%
{{\bfseries \tablename\ \thetable{} -- continued from previous page}} \\
\midrule  \textbf{Classification Task} & \textbf{Receiver Loc.} & \multicolumn{5}{c}{\textbf{Signal}} & \multicolumn{5}{|c|}{\textbf{Threema}} &  \multicolumn{5}{c}{\textbf{Whatsapp}}\\ 
&& 1&2&3&4&5&1&2&3&4&5&1&2&3&4&5\\
\midrule
\endhead

\midrule \multicolumn{17}{r}{{Continued on next page}} \\ \midrule
\endfoot

\hline \hline
\endlastfoot

\multicolumn{17}{l}{\emph{Two cellular locations}}\\
\midrule
2mloc1-DE-2ALL&Overall Acc.&0.55&0.6&0.64&0.66&0.73&0.74&0.87&0.85&0.85&0.85&0.79&0.83&0.85&0.86&0.88\\
2mloc1-DE-2ALL&m-DE-A&0.69&0.9&0.81&0.82&0.75&0.56&0.93&0.9&0.91&0.91&0.59&0.73&0.78&0.82&0.89\\
2mloc1-DE-2ALL&m-DE-B&0.42&0.31&0.46&0.5&0.71&0.93&0.81&0.79&0.8&0.79&1.0&0.93&0.91&0.89&0.88\\
\midrule
\multicolumn{17}{l}{\emph{Two cellular locations and any WiFi location}}\\
\midrule 
2mlocw1-DE-2ALL&Overall Acc.&0.5&0.57&0.61&0.66&0.75&0.55&0.71&0.63&0.68&0.68&0.61&0.73&0.74&0.77&0.81\\
2mlocw1-DE-2ALL&m-DE-A&0.72&0.65&0.69&0.63&0.74&0.57&0.73&0.68&0.75&0.66&0.58&0.68&0.74&0.78&0.82\\
2mlocw1-DE-2ALL&m-DE-B&0.06&0.29&0.37&0.51&0.66&0.61&0.77&0.66&0.74&0.82&0.64&0.72&0.7&0.71&0.77\\
\end{longtable}
\end{center}

\end{document}